\documentclass[12pt,journal,compsoc]{IEEEtran}

\usepackage{graphicx}
\usepackage[table]{xcolor}
\usepackage{multirow}
\usepackage{siunitx}
\usepackage{subcaption}
\usepackage[most]{tcolorbox}
\usepackage[export]{adjustbox}
\usepackage{array}
\usepackage{rotfloat}
\usepackage[T1]{fontenc}
\usepackage{lscape}
\usepackage{amssymb}
\usepackage{framed}
\usepackage{lineno}
\usepackage[utf8]{inputenc}
\usepackage{longtable}
\usepackage{supertabular,booktabs}

\ifCLASSOPTIONcompsoc
\else
\fi

\ifCLASSINFOpdf
\else
\fi

\begin{document}
%
\title{Software Module Clustering: An In-Depth Literature Analysis}

\author{Qusay I. Sarhan, 
        Bestoun S. Ahmed, 
        Miroslav Bures, and
        Kamal Z. Zamli 
\IEEEcompsocitemizethanks{\IEEEcompsocthanksitem Q. Sarhan is with the Department of Computer Science, University of Duhok, Duhok, Kurdistan Region, Iraq\hfil\break
E-mail: qusay.sarhan@uod.ac

\IEEEcompsocthanksitem B. Ahmed is with the Department of Mathematics and Computer Science, Karlstad University, 651 88 Karlstad, Sweden\hfil\break
E-mail: bestoun@kau.se

\IEEEcompsocthanksitem M. Bures is with the Department of Computer Science, Faculty of Electrical Engineering, Czech Technical University in Prague, Karlovo nam. 13, 121 35 Praha 2, Czech Republic\hfil\break
E-mail: buresm3@fel.cvut.cz

\IEEEcompsocthanksitem K. Zamli is with the IBM Center of Excellence, Faculty of Computer Systems and Software Engineering, University Malaysia Pahang, 26 300 Gambang, Kuantan, Pahang, Malaysia
\hfil\break
E-mail: kamalz@ump.edu.my
}

}

\markboth{Journal of \LaTeX\ Class Files,~Vol.~11, No.~4, December~2012}%
{Shell \MakeLowercase{\textit{et al.}}: Bare Advanced Demo of IEEEtran.cls for Journals}
%

\IEEEtitleabstractindextext{%
\begin{abstract}
Software module clustering is an unsupervised learning method used to cluster software entities (e.g., classes, modules, or files) with similar features. The obtained clusters may be used to study, analyze, and understand the software entities' structure and behavior. Implementing software module clustering with optimal results is challenging. Accordingly, researchers have addressed many aspects of software module clustering in the past decade. Thus, it is essential to present the research evidence that has been published in this area. In this study, 143 research papers from well-known literature databases that examined software module clustering were reviewed to extract useful data. The obtained data were then used to answer several research questions regarding state-of-the-art clustering approaches, applications of clustering in software engineering, clustering processes, clustering algorithms, and evaluation methods. Several research gaps and challenges in software module clustering are discussed in this paper to provide a useful reference for researchers in this field.
\end{abstract}

}

\maketitle

\IEEEdisplaynontitleabstractindextext

\IEEEpeerreviewmaketitle

\section{Introduction}
\IEEEPARstart{C}{lustering} (also called cluster analysis) is an unsupervised data mining technique that groups a set of data points into several clusters \cite{ADOLFSSON201913}. When several points fall inside a cluster, they are similar in some features. Measurement of the similarity and dissimilarity relies on the extent to which the data points share the same features. Clustering has been employed in many important fields of study and applications, including software engineering, information retrieval, machine learning, pattern recognition, and statistics \cite{ref2-2009}.

In the context of software engineering, software clustering is being defined as the process of decomposing large software systems into smaller, manageable, meaningful (highly cohesive), independent (loosely coupled), and feature-oriented (share common features) subsystems \cite{ref11-2008, ref1-2011}. These subsystems may contain entities/artifacts (e.g., classes, modules, or files) of similar features. A software module clustering approach that accomplishes this task can have a substantial impact and practical benefits, especially for developers working on legacy systems with documentation that is outdated or nonexistent. Clustering in software engineering can be used in many applications, such as architecture recovery \cite{ref4-2009}, code clone detection \cite{ref10-2015}, and poor design detection \cite{ref5-2012}.

With several algorithms and studies published in the literature, software module clustering has become an active research area. Although many module clustering approaches have been proposed and applied, they have difficulties meeting the current development and advancement needs in software and its various applications. For example, there is a lack of experimental studies on clustering a system developed using more than one programming language \cite {ref10-2019} or a system that performs some of its tasks by invoking ready-made web services available on the Internet/network \cite{webservices2007}, \cite{webservices2017}.

This paper presents a comprehensive systematic literature study to structure and categorize the state-of-the-art research evidence related to software module clustering during the past decade. The study defines several research questions (RQs) that cover many aspects of the field and then identifies relevant papers and their results. It concludes by discussing future research opportunities in the area. In this process, a systematic method is used to collect and analyze the related published research.

The remainder of this paper is structured as follows: Section \ref{MotivationRelatedWork} presents the motivation and an overview of related work. Section \ref{ResearchMethod} describes in detail the research methodology used to conduct this study. 
Section \ref{Results} presents the results and outcomes. Section \ref{Threats} discusses the issues of validity. Finally,  Section \ref{Conclusion} presents the conclusions of the study.



\section{Motivation and Related Work}\label{MotivationRelatedWork}

Software module clustering is an important topic of research in software engineering. Although it started in the 1990s, software module clustering research has experienced more momentum and attention in the past decade.  This momentum and attention are reflected in the dramatic increase in the number of publication. Among the many factors leading to this increased attention, there have been two leading factors in the past decade. First is the dramatic increase in software application size due to the newly added functionalities and features they provide. This, in turn, has led to an increase in the number of modules for these applications. Here, software module clustering is a good approach to manage and maintain this kind of application. Second, the advancement of artificial intelligence (AI) methods (such as data mining, clustering, optimization, and machine learning methods) in the past decade  has played a substantial role in increasing the research activity related to software module clustering.

Although it has been an active research area since the 1990s, a systematic literature study exploring research on software module clustering in the past decade is lacking. A few survey/review studies on software module clustering have been found in the literature. Shtern and Tzerpos \cite{other1} provided an overview of different software clustering methods and their applications in software engineering. The study highlighted some approaches for the evaluation of software clustering results. It also presented some research challenges to be addressed to improve software clustering results. In \cite{other10}, the basic concepts and necessity of software module clustering were presented briefly. The authors also described different metaheuristic search techniques that have been applied to the software module clustering problem in the maintenance phase of the software development life cycle. In \cite{other11}, the authors presented different search-based approaches to software clustering that have been classified into several categories (mono-objective, multiobjective, and many-objective) based on the number of clustering quality criteria. Furthermore, the advantages and disadvantages of each category are presented briefly. In \cite{other12} and its extended version \cite{other13}, the authors described search-based optimization techniques and their applications in different software engineering domains. They briefly introduced software modularization and refactoring as clustering problems that can be addressed using several search algorithms. Additionally, they presented some research challenges with search algorithms, including determining suitable stopping criteria to terminate the search and issues related to visualizing the search results. The authors in \cite{other14} also dedicated parts of their study to performing software modularization and refactoring using search-based optimization techniques. They briefly introduced a number of algorithms in this respect, including NSGA-II and PCA-NSGA-II. Additionally, a number of evaluation metrics, such as coupling, cohesion, and MQ (modularization quality) have been mentioned.

The aforementioned studies were not focused on conducting a dedicated literature analysis of software module clustering: some aspects of software module clustering are covered as part of other related topics. By contrast, our paper presents an in-depth and systematic analysis with a detailed research methodology to examine different software module clustering aspects, such as applications, algorithms, tools, target systems, evaluation methods, and possible research gaps.
\section{Research Methodology}\label{ResearchMethod}

\begin{figure}
\centering
\includegraphics[width=3 in, height=8cm]                {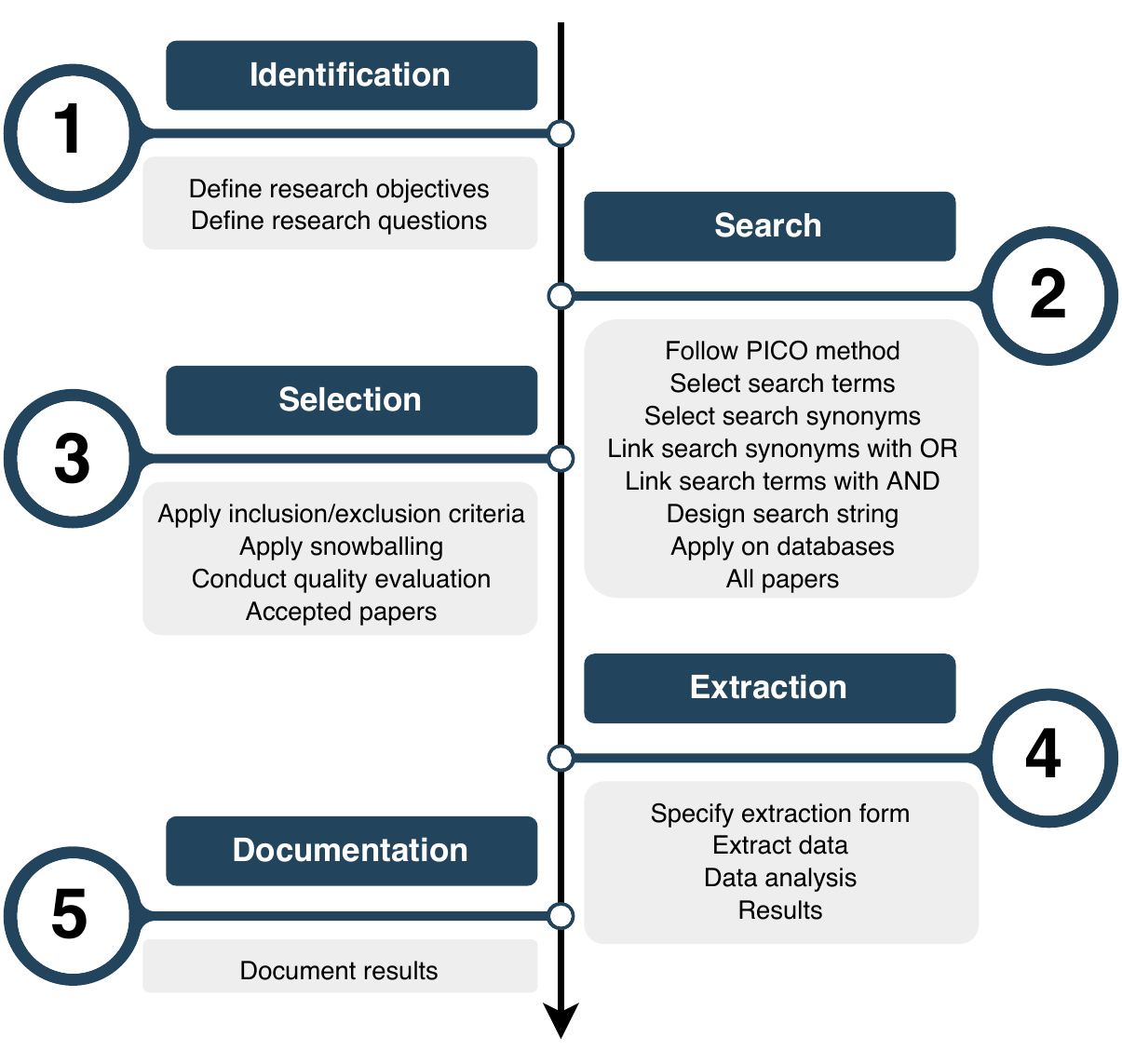}
\caption {Stages of the used  literature analysis study.}
\label{Fig:1}
\end{figure}

This study was based on the guidelines of conducting systematic mapping studies provided by \cite{other2}, the guidelines for conducting literature review studies provided by \cite{other3}, and other studies. These studies aimed to investigate software engineering in a particular context other than software module clustering (e.g., \cite{other4,other5,other6}). Figure \ref{Fig:1} shows the five-stage systematic process used in this research.

The first stage is defining the research scope and questions. The main research problem of the study was identified, and several questions were formed to address the problem. The second stage is conducting the search process in which a search strategy to select the primarily published papers was specified. The third stage is screening the published papers obtained from the previous stage via several filtering methods. The fourth stage is data extraction, where the selected papers are carefully analyzed, and useful data are extracted to answer the questions defined in this study. Finally, the fifth stage is the results reporting. The following subsections elaborate on the aforementioned stages in detail.


\subsection{Identification of the Research Need and Scope}

\subsubsection{Research Objectives}
This study aims to provide a comprehensive analysis of publications on software module clustering in the past decade by identifying, examining, and categorizing state-of-the-art contributions. Only papers from the past decade were considered to keep the review focused on recent works on the topic. Notable changes in software development have occurred over the past decade. Software systems migrated from simple architectures, such as monolithic and two-tiered architectures to multitier (also called n-tier or multilayered) architectures. Additionally, development approaches changed from structured to object-oriented to service-oriented. In other words, software development continues to change. Software developers use vast amounts of source code to understand large and complex systems before making changes, which is not feasible. Thus, software clustering has been employed to provide automated assistance to recover the abstract structure of systems and understand them.

\subsubsection{Research Questions}
To achieve the aim of this study and reflect its objectives, several RQs related to the topic were defined. Each defined RQ addresses a specific aspect of software module clustering.

\begin{itemize}
\item RQ1. What is the number and distribution of publications on software module clustering over the last decade?
\item RQ2. Which researchers, organizations, and countries are active in software module clustering research?
\item RQ3. What are software module clustering applications and how these applications are distributed?
\item RQ4. What is the standard process of software module clustering?
\item RQ5. What are the software systems used as targets for the experiments of software module clustering?
\item RQ6. What are the current factbase sources, their types, their forms, and extraction tools used in software module clustering?
\item RQ7. What are the most used similarity measures in the software module clustering?
\item RQ8. What are the most used algorithms, their types, and standard stop conditions?
\item RQ9. What are the most used tools for visualizing clustering results?
\item RQ10. What are the metrics used to evaluate clustering results and the current approaches to obtaining the gold/expert decomposition?
\item RQ11. What are the potential future research directions on software module clustering?
\end{itemize}


\subsection{Search Strategy }

\subsubsection{Literature Sources} 

In this study, five standard online databases were selected as sources that index the literature of software engineering and computer science \cite{other2}. These sources are presented in Table \ref{Tab:2}.

\begin{table}[h]
    \centering
\captionof{table}{The used database sources to explore the literature.} 
\label{Tab:2}
\begin{tabular}{ |l|l| } 
 \hline
  Source & URL  \\
 \hline
  IEEE Xplore & https://ieeexplore.ieee.org/\\ 
 \hline
  Elsevier ScienceDirect & https://sciencedirect.com/\\ 
  \hline
  ACM Digital Library & https://dl.acm.org/ \\
  \hline
  Scopus & https://scopus.com/\\
  \hline
  SpringerLink & https://link.springer.com/\\ 
 \hline
\end{tabular}
\end{table}

\subsubsection{Search String }

The population, intervention, comparison, and outcomes (PICO) method \cite{other3} was employed to identify related studies. Here, population (P) refers to the applied area of clustering, intervention (I) refers to the process or procedure used to solve the clustering problem and outcomes (O) refers to the outcomes of the work and research of clustering in software engineering. Comparison (C) is not considered in the keywords, as this study is a general analysis of clustering in software engineering. Table \ref{Tab:3} presents the keywords associated with each part of the method in the final search string. Notably, the search terms used in the final search string were linked with one another based on the steps described in \cite{other9}. Here, the "OR" Boolean operator was used to link synonyms or related search terms to the topic of this study and the "AND" Boolean operator was employed to link the main search terms.

\begin{table*}
    \centering
\captionof{table}{PICO related keywords.} 
\label{Tab:3}
\begin{tabular}{ |l|p{13cm}| } 
 \hline
PICO method & Keywords used  \\
 \hline
Population (P) & 
(software OR project OR system OR application OR program OR module OR component OR service OR source code OR package OR file OR function OR class)  \\ 
 \hline
 Intervention (I) & (clustering OR cluster analysis OR partitioning OR grouping OR splitting OR structuring OR modularizing OR constructing OR composing OR categorizing OR classifying) \\ 
  \hline
 Comparison (C) & No comparisons considered in the keywords \\
  \hline
 Outcomes (O) & (methodology OR algorithm OR technique OR approach OR method OR tool OR improvement OR evaluation OR similarity measurement OR application OR metric OR problem OR challenge OR limitation) \\
 \hline
\end{tabular}
\end{table*}


Different combinations of search strings tried to construct the final one since the term "clustering" is used in other research areas such as data mining, image processing, and statistics. The final search string was the one that meets the following two criteria.

\begin{itemize}
\item The search string that returns the most relevant studies.
\item The search string that returns the maximum number of the identified pilot set. For this study, a pilot set of 25 papers has been selected based on our experience and initial research review. 
\end{itemize}

As an example, the search string Try 2 in Table \ref{Tab:searchstrings} was excluded because the returned result is incomplete compared to the search string Try 5.

\begin{table*}
    \centering
\captionof{table}{Search strings piloted on IEEE Xplore.} 
\label{Tab:searchstrings}
\begin{tabular}{ |l|p{10cm}|l|l| } 
 \hline
  \# Try & Search string & \# Results & \# Missing studies  \\
 \hline
 Try 1  & ((software) AND (clustering) AND (problem)) & 355  & 15\\ 
 \hline
  Try 2 & ((software OR module OR component) AND (clustering OR cluster analysis) AND (algorithm OR technique OR approach OR problem)) & 450  & 11\\ 
  \hline
  Try 3 & ((software OR system  OR program OR module OR component OR source code OR package OR file) AND (clustering OR cluster analysis OR partitioning OR grouping OR splitting OR structuring OR categorizing OR classifying) AND (methodology OR algorithm OR technique OR approach OR method OR tool OR improvement OR problem OR challenge OR limitation)) & 631  & 8\\
  \hline
  Try 4 & ((software OR system  OR program OR module OR component OR source code OR package OR file) AND (clustering OR cluster analysis OR partitioning OR grouping OR splitting OR structuring OR categorizing OR classifying) AND (methodology OR algorithm OR technique OR approach OR method OR tool OR improvement OR measurement OR application OR  problem OR challenge OR limitation)) & 811 & 3\\
  \hline
  Try 5 & ((software OR project OR system OR application OR program OR module OR component OR service OR source code OR package OR file OR function OR class) AND (clustering OR cluster analysis OR partitioning OR grouping OR splitting OR structuring OR modularizing OR constructing OR composing OR categorizing OR classifying) AND (methodology OR algorithm OR technique OR approach OR method OR tool OR improvement OR evaluation OR similarity measurement OR application OR metric OR problem OR challenge OR limitation)) & 905  & 0\\ 
 \hline
\end{tabular}
\end{table*}

\subsection{Paper Selection}

\subsubsection{Inclusion/Exclusion Criteria}

To decide whether a paper is relevant to the scope of this research, a set of criteria, which are presented below, for inclusion and exclusion  was considered.

\begin{itemize}

\item \textbf{Inclusion criteria are:}

\begin{itemize}

\item Papers published online from 2008-2019.  
Papers related directly to software module clustering. This is ensured by reading the title of each obtained paper. However, the abstract or full-text reading has been also applied when the title reading was not enough. This criterion filtered most of the papers out.

\end{itemize}

\

\item \textbf{Exclusion criteria are:}

\begin{itemize}
\item Papers not published in English are excluded since English is the prevalent language used in the
scientific peer-reviewing global community.
\item Papers without accessible full text.
\item Papers not formally peer-reviewed (gray literature and books).
\item Papers not published electronically.
\item The duplicated papers were excluded from the list. Authors sometimes publish expanded versions of their conference papers to journal venues. Such papers share most of the material and considering them both would affect the quality of this study. To overcome this issue, duplicated papers are identified by comparing paper titles, abstracts, and contents. When the duplication is confirmed, the least recent publication is removed.

\item Master and Ph.D. dissertations are excluded because the content of such publications is eventually presented in peer-reviewed venues, which have already been considered in our study.
\item Papers that are published as surveys are filtered out because they do not actually bring new technical contributions to software module clustering.
\end{itemize}
\end{itemize}


\subsubsection{Snowballing}
The snowballing \cite{other7} search method was applied to the remaining papers to reduce the possibility of missing critical related papers. In this method, each research paper's list of references is examined in terms of the previously applied inclusion/exclusion criteria. The process was then recursively applied to newly identified papers.

\subsubsection{Quality Evaluation}
For quality evaluation, each paper must be evaluated to determine whether sufficient information can be extracted from it. Papers that did not provide answers to the following two questions were excluded:

\begin{itemize}
\item Is the process of software module clustering described in detail? 

\item Are there experimental results and evaluation for the software module clustering process?

\end{itemize}

\

\begin{figure}
\centering
\includegraphics[width=3.4 in]                {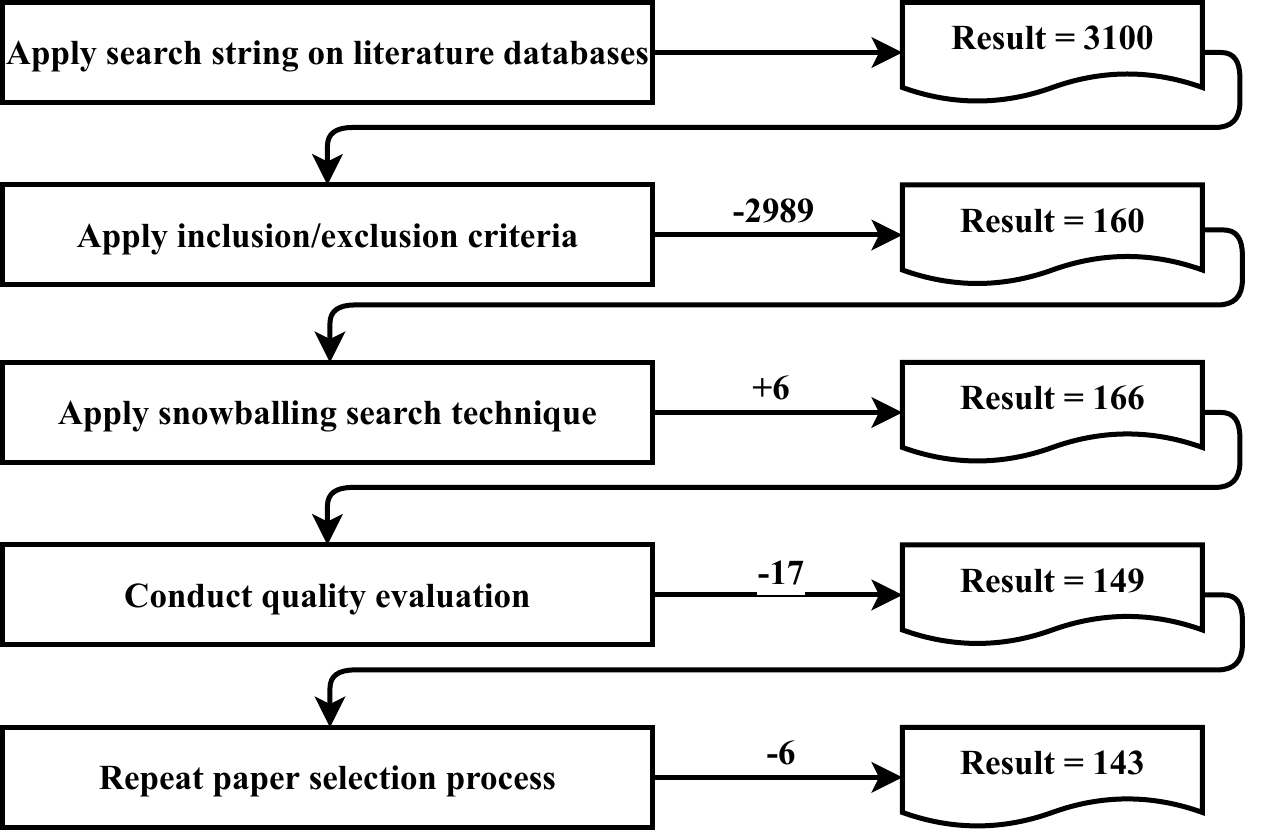}
\caption{Results of the paper selection process.}
\label{Fig:19}
\end{figure}

The number of included and excluded papers at each stage of the paper selection process is shown in Figure \ref{Fig:19}. Besides, all the identified papers and their references, full names, and publication years are listed in Table \ref{Tab:11}.

\subsection{Data Extraction}
In this phase, the data of the considered studies were extracted and analyzed to answer the defined RQs. The data obtained in this phase were stored in an Excel file with different fields created for this purpose. Each data extraction field has a data item and a value, as presented in Table \ref{Tab:4}. Notably, a reliable data extraction method was followed in this stage: the data were extracted first by the first author and then double-checked by the other authors separately.

\begin{table*}
    \centering
\captionof{table}{Data extraction form.} 
\label{Tab:4}
\begin{tabular}{ |l|l|l| } 
 \hline
 Data  Item &   Value & Relevant RQ\\
 \hline
 ID &  Integer paper ID number  & None\\ 
 \hline
  Title &  Paper title & None\\
  \hline
  Year &   Paper publication year & RQ1\\
 \hline
Type &   Paper publication type & RQ1 \\
 \hline
 Venue &  Publication venue name& RQ1\\
 \hline
 Author Name &   Name of the author(s) & RQ2\\
  \hline
 Author Organization &   Name of the organization for each participated author/co-author & RQ2\\
  \hline
 Author Country &   Name of the country for each participated author/co-author & RQ2\\
 \hline
 Clustering Applications &   Applications of software module clustering & RQ3\\
 \hline
 Clustering Process &   The standard steps of software module clustering & RQ4\\
 \hline
 Target Systems &   The software systems used in the experimental testing & RQ5\\
 \hline
 Factbase Extraction &   The sources, fact types, and tools used for factbase extraction & RQ6\\
 \hline
Similarity Measures &  The similarity measures used in the clustering process & RQ7\\
 \hline
 Clustering Algorithms &  The algorithms used in clustering and their types & RQ8\\
 \hline
 Visualization Tools &   The tools used for visualizing clustering results & RQ9\\
 \hline
 Evaluation Metrics &   The metrics used for clustering results evaluation & RQ10\\
 \hline
 Future Research &   The possible future research directions & RQ11\\
 \hline
\end{tabular}
\end{table*}


\section{Results}\label{Results}
The selected papers were carefully analyzed to answer the RQs. Here, a short title is used to represent each RQ. The following subsections present and discuss the results based on each RQ.


\subsection{Distribution of Publications (RQ1)}

\subsubsection{Publication Frequency}

\begin{figure}
\centering
\includegraphics[width=3.4 in]                {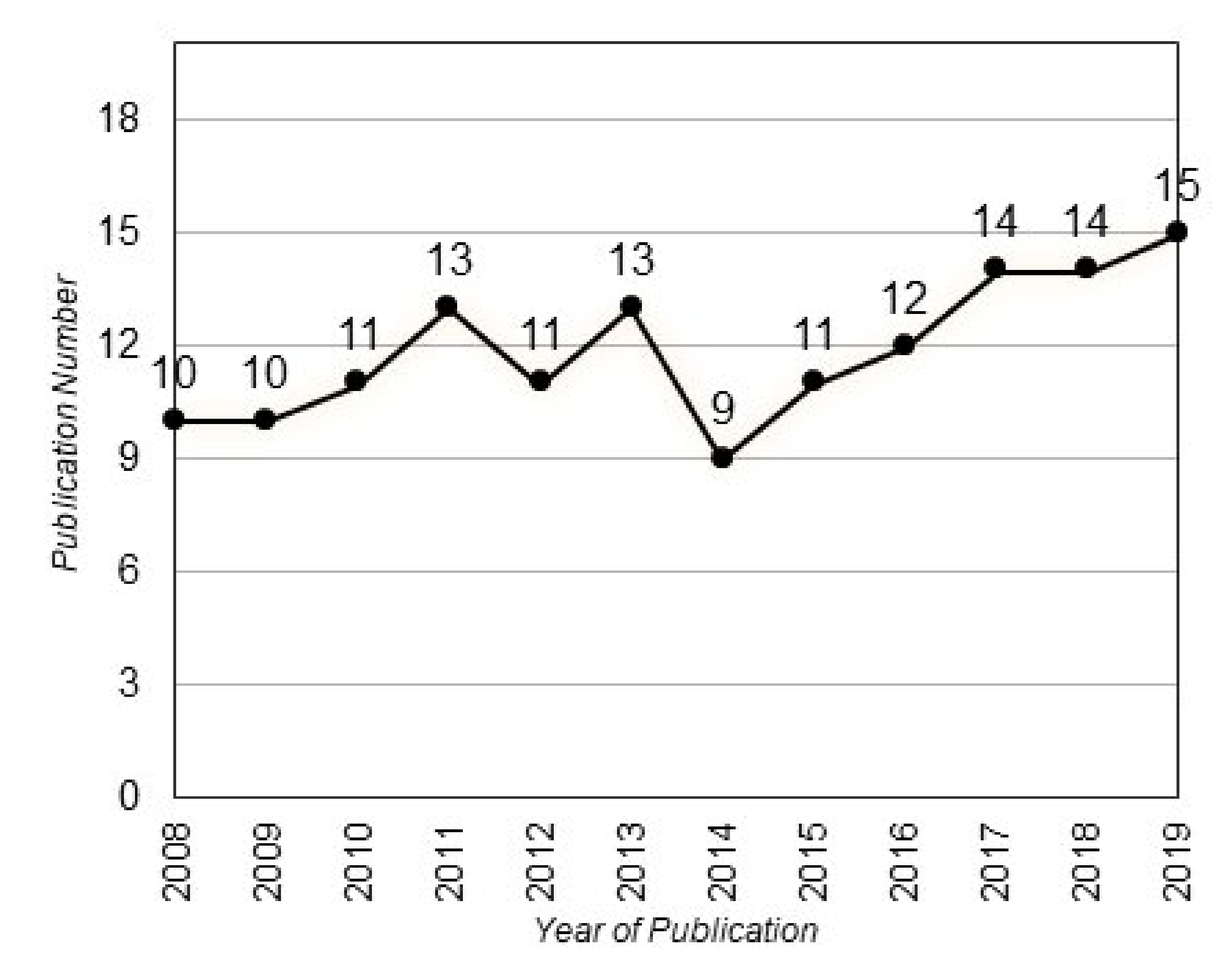}
\caption{Publication per year.}
\label{Fig:2}
\end{figure}

As previously mentioned, 143 papers published over the past decade (2008-2019) were included in this study. The selected papers were analyzed to determine their frequency and evolution, as shown in Figure \ref{Fig:2}.

The figure also shows that the average number of publications per year is approximately 12. Additionally, interest in the topic has increased in the past three years, which indicates the successful application of module clustering to solve problems in software engineering along with the exponential growth of software applications in number, size, and complexity. One reason for this growth is that software has no limited lifetime: software code is constantly changed to meet user needs. Thus, developers are always in need of tools and approaches to ease the process of software maintenance. Software module clustering helps considerably in this context. Another valid reason is that software development is constantly changing with the development of new technologies. In the past few years, AI and the Internet of Things (IoT) have become trending technologies \cite{trend1} impacting software development. The complexity of software systems based on such technologies is increasing dramatically as advanced features are employed. As a result, many legacy systems have been transformed to cope with this change. Understanding those systems before integrating them with new technologies requires an in-depth analysis, which may be achieved by employing software clustering.

\subsubsection{Publication Venue}

\begin{figure}
\centering
\includegraphics[width=3.5 in]                {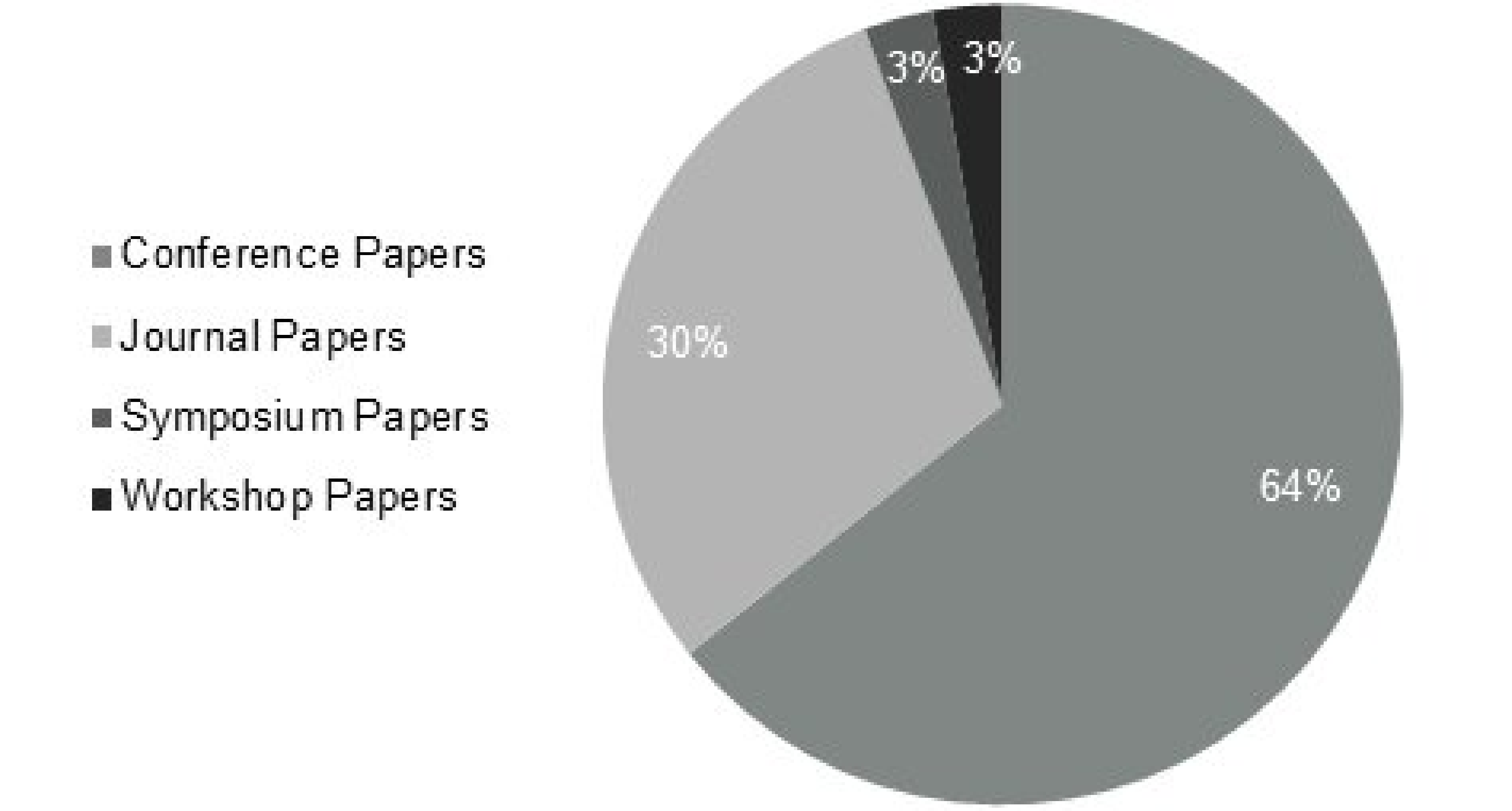}
\caption{Publication ratio per each venue.}
\label{Fig:3}
\end{figure}

\begin{figure}
\centering
\includegraphics[width=3.8 in]                {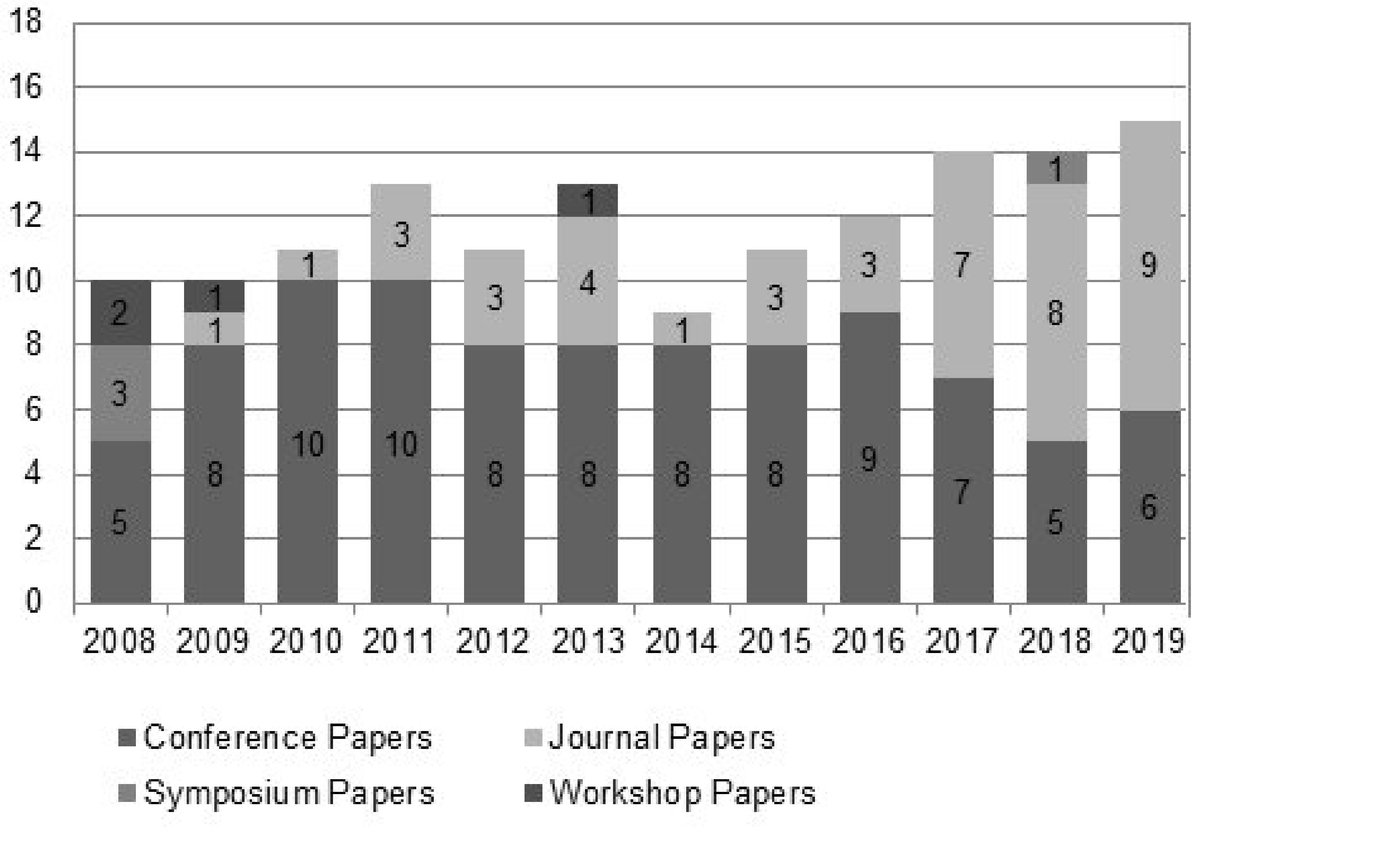}
\caption{Publication number per each venue.}
\label{Fig:4}
\end{figure}

Figure \ref{Fig:3} shows that the considered papers, namely, 92 are conference papers, 43 are journal papers, 4 are symposium papers, and 4 are workshop papers were published in various venues. This result also shows that only 30\% of the considered papers have reached the maturity of a journal publication, indicating that software module clustering is a very young or even immature research area \cite{young1}, \cite{young2}. Moreover, few conference papers were published as book chapters. For these papers, consideration was given to their original venues, that is conferences. Figure \ref{Fig:4} shows the publication number per year by venue type.

\begin{figure}
\centering
\includegraphics[width=4in , height=3in]                {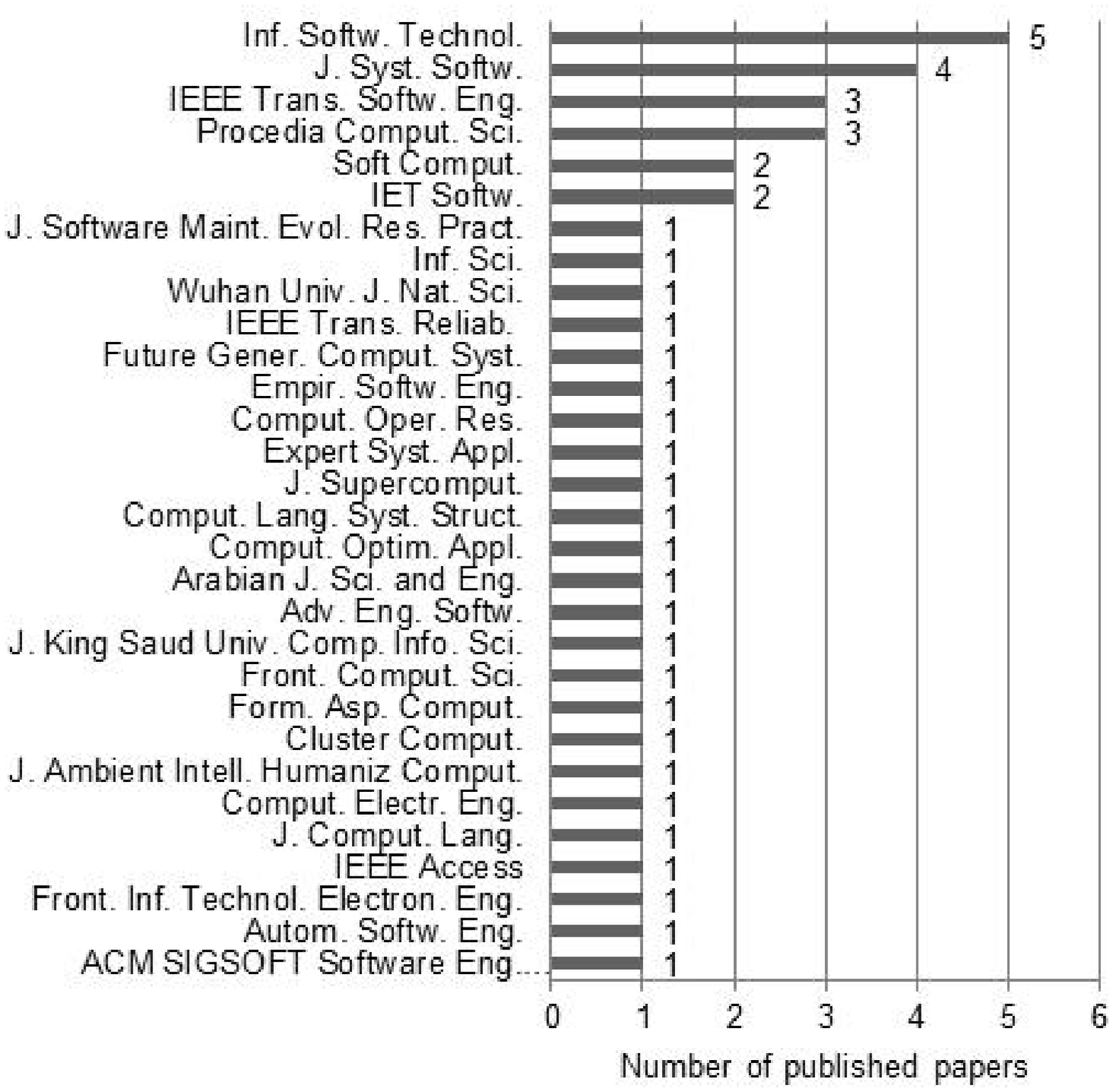}
\caption{Number of published papers vs. journal name.}
\label{Fig:5}
\end{figure}

The most active and top journals, conferences, symposiums, and workshop venues that publish papers on software module clustering can be determined by analyzing the publications. Abbreviations are used in this paper instead of full names. Figure \ref{Fig:5} shows the active journals in which the considered studies were published. The full names of the journals are presented in Table \ref{Tab:12}. The figure also shows that the most active and top journals are  "Inf. Softw. Technol.", "J. Syst. Softw.", "IEEE Trans. Softw. Eng.", "Procedia Comput. Sci.", "Soft Comput.", and "IET Softw.". Notably, 44\% of the journal papers were published in the top six journals, whereas the other 56\% were published in individual journals.

\begin{figure}
\centering
\includegraphics[width=3.4 in, height=5cm]                {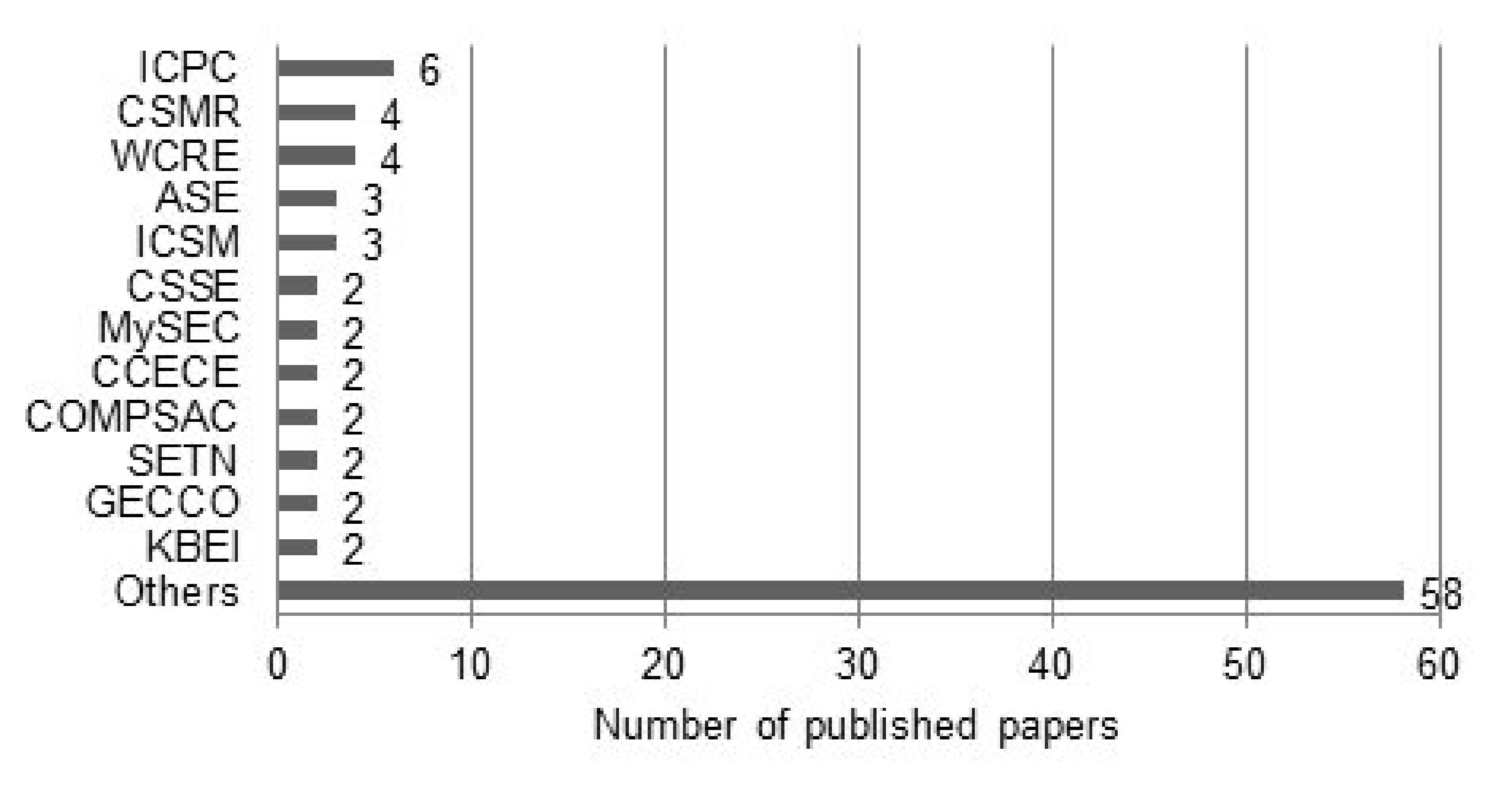}
\caption{Number of published papers vs. conference name.}
\label{Fig:6}
\end{figure}

Figure \ref{Fig:6} shows the active conferences that published papers in software module clustering. The full names of the conferences can be found in Table \ref{Tab:13}. The most active and top conferences are the International Conference on Program Comprehension (ICPC), Conference on Software Maintenance and Reengineering (CSMR), and Working Conference on Reverse Engineering (WCRE). Notably, approximately 15\% of the conference papers were published at these top three conferences. If other conferences that publish two or three papers are considered, then approximately 37\% of the conference papers were published by annual conferences. The largest number, approximately 63\%, of the published conference papers were at individual conferences, represented as "Others" in Figure \ref{Fig:6}.

\begin{figure}
\centering
\includegraphics[width=3.4 in]                {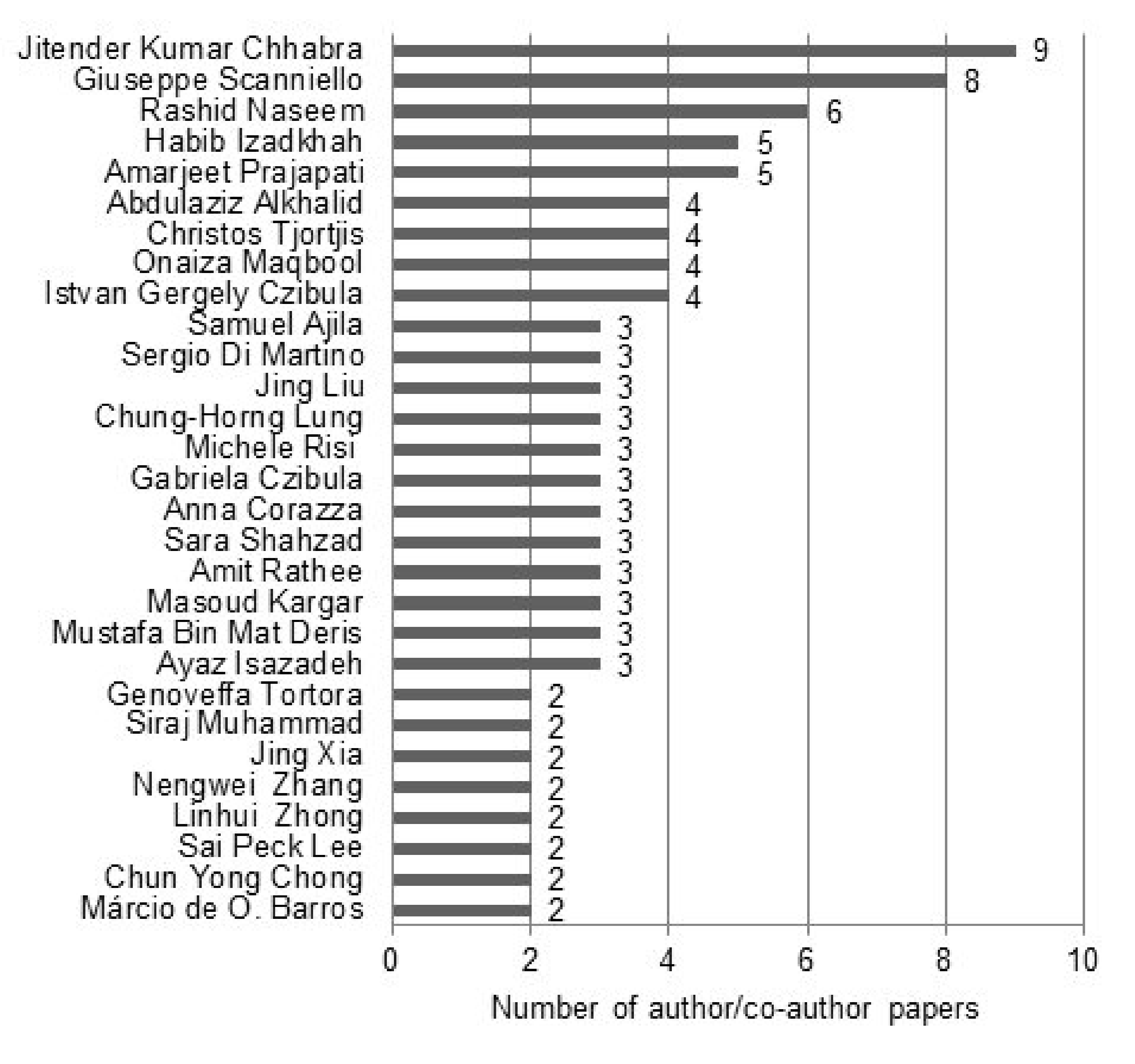}
\caption{Active researchers based on the published papers.}
\label{Fig:7}
\end{figure}


\subsection{Active Researchers, Organizations, and Countries (RQ2)}

Many researchers are interested and involved in software module clustering research. However, the most active researchers are those who have more than two published papers either as a main author or coauthor. The ranking of active researchers is shown in Figure \ref{Fig:7}. The ranking shows that "Jitender Kumar Chhabra" and "Giuseppe Scanniello" are the top two researchers in this field, with nine and eight published papers, respectively. These authors participated in approximately 12\% (17/143) of all publications.

Table \ref{Tab:5} shows the ranking of the active countries and organizations, including the name of country, organization, participating researchers, reference to the published papers, and the total number of papers.

\begin{table*}
    \centering
      
\captionof{table}{Countries, organizations, and researchers active in researching software module clustering.} 
\label{Tab:5}
\begin{tabular}{ |p{1.9cm}|p{5cm}|p{5cm}|p{3cm}|l| } 
 \hline
 Country & Organization &   Author(s) & Published Papers & Total\\

 \hline
India  &National Institute of Technology Kurukshetra &    Jitender Kumar Chhabra, Amarjeet Prajapati, and Amit Rathee &    
  
  \cite{ref4-2011}, \cite{ref2-2014}, \cite{ref2-2017}, \cite{ref3-2017}, \cite{ref6-2017}, \cite{ref7-2017}, \cite{ref10-2017}, \cite{ref10-2018}, \cite{ref14-2018}, \cite{ref15-2019}
  &    10\\
  
  \hline
Italy &University of Basilicata &    Giuseppe Scanniello &    
 \cite{ref3-2010}, \cite{ref13-2010},  \cite{ref3-2011}, \cite{ref6-2011}, \cite{ref8-2011}, \cite{ref11-2012}, \cite{ref2-2013},  \cite{ref12-2016} 
 &  8  \\

 \hline
Iran &University of Tabriz &    Habib Izadkhah, Hamid Masoud, and Ayaz Isazadeh&    
 \cite{ref6-2017}, \cite{ref2-2019}, \cite{ref11-2014},  \cite{ref9-2019},  \cite{}, \cite{ref13-2019}, \cite{ref14-2019}
 &    7 \\  
  
  \hline
  
Pakistan &City University of Science and Information Technology &    Rashid Naseem &    
 
 \cite{ref5-2012}, \cite{ref4-2013}, \cite{ref1-2016}, \cite{ref6-2019}, \cite{ref11-2013}  
 &    5 \\ 
 
  \hline
 Pakistan&Quaid-i-Azam University &    Onaiza Maqbool, Adeel Ahmed, and A.Q. Abbasi    &    
 \cite{ref5-2012}, \cite{ref4-2013}, \cite{ref6-2019}, \cite{ref8-2012},  \cite{ref13-2017}, 
 &    5 \\
 
 \hline
Romania &Babeș-Bolyai University &    Istvan Gergely Czibula and Gabriela Czibula &     
  
  \cite{ref11-2008}, \cite{ref6-2008},   \cite{ref6-2010}, \cite{ref11-2017}
  
  &    4 \\ 
  \hline
 
 Canada&Carleton University    &    Samuel Ajila, Chung-Horng Lung, and Abdulaziz Alkhalid &        
 \cite{ref9-2013}, \cite{ref10-2013}, \cite{ref3-2015}
 
 & 3 \\ 
 \hline
Italy &University of Naples Federico II &    Anna Corazza and Sergio Di Martino    & 
 
 \cite{ref3-2010}, \cite{ref8-2011}, \cite{ref12-2016}
 &    3 \\ 
 
 \hline
  Malaysia & Universiti Tun Hussein Onn Malaysia &     Mustafa Mat Deris and Rashid Naseem &    
 \cite{ref1-2016}, \cite{ref6-2019} , \cite{ref13-2017}
 &    3 \\

 \hline
 China&Xidian University    & Jing Liu &     
 \cite{ref2-2016}, \cite{ref1-2017}, \cite{ref3-2018}
 & 3 \\

 \hline
 Italy&University of Salerno &    Michele Risi  and Genoveffa Tortora  &    
 \cite{ref3-2011}, \cite{ref11-2012}, \cite{ref4-2010}
 &    3 \\
 
 \hline
Iran   &Islamic Azad University &     Masoud Kargar &    
 \cite{ref6-2017}, \cite{}, \cite{ref14-2019} 
 &    3 \\
 
 \hline
  Pakistan &University of Peshawar &    Sara Shahzad &    
 \cite{ref6-2019} , \cite{ref11-2013}, \cite{ref13-2017} 
 &    3 \\
 
 \hline
Saudi Arabia
 &King Abdullah University of Science and Technology    & Abdulaziz Alkhalid &    
 \cite{ref12-2010}, \cite{ref12-2011}
 &    2 \\ 
 
  \hline
Pakistan &Shaheed Benazir Bhutto University &    Siraj Muhammad    &    
 \cite{ref4-2013}, \cite{ref8-2012} 
 &    2 \\
 
 \hline
Brazil &UNIRIO - Universidade Federal do Estado do Rio de Janeiro &    Márcio de O. Barros    &    
 \cite{ref9-2012}, \cite{ref4-2017}
 &    2 \\
 
 \hline
England &University of Manchester &    Christos Tjortjis &     
 
 \cite{ref2-2009}, \cite{ref7-2008}, 
 &    2 \\ 
 \hline
Greece &International Hellenic University &    Christos Tjortjis &    
 \cite{ref6-2016}, \cite{ref10-2014}
 &    2 \\
 
  \hline
China   & Jiangxi Normal Universit &     Linhui Zhong, Nengwei Zhang, and Jing Xia &    
 \cite{ref4-2016}, \cite{ref14-2016}
 &    2 \\
 
   \hline
Malaysia   & University of Malaya &     Chun Yong Chong and Sai Peck Lee &    
 \cite{ref6-2013}, \cite{ref12-2015}
 &    2 \\
 
 \hline
\end{tabular}
\end{table*}

\begin{figure}
\centering
\includegraphics[width=3.4 in, height=5cm]                {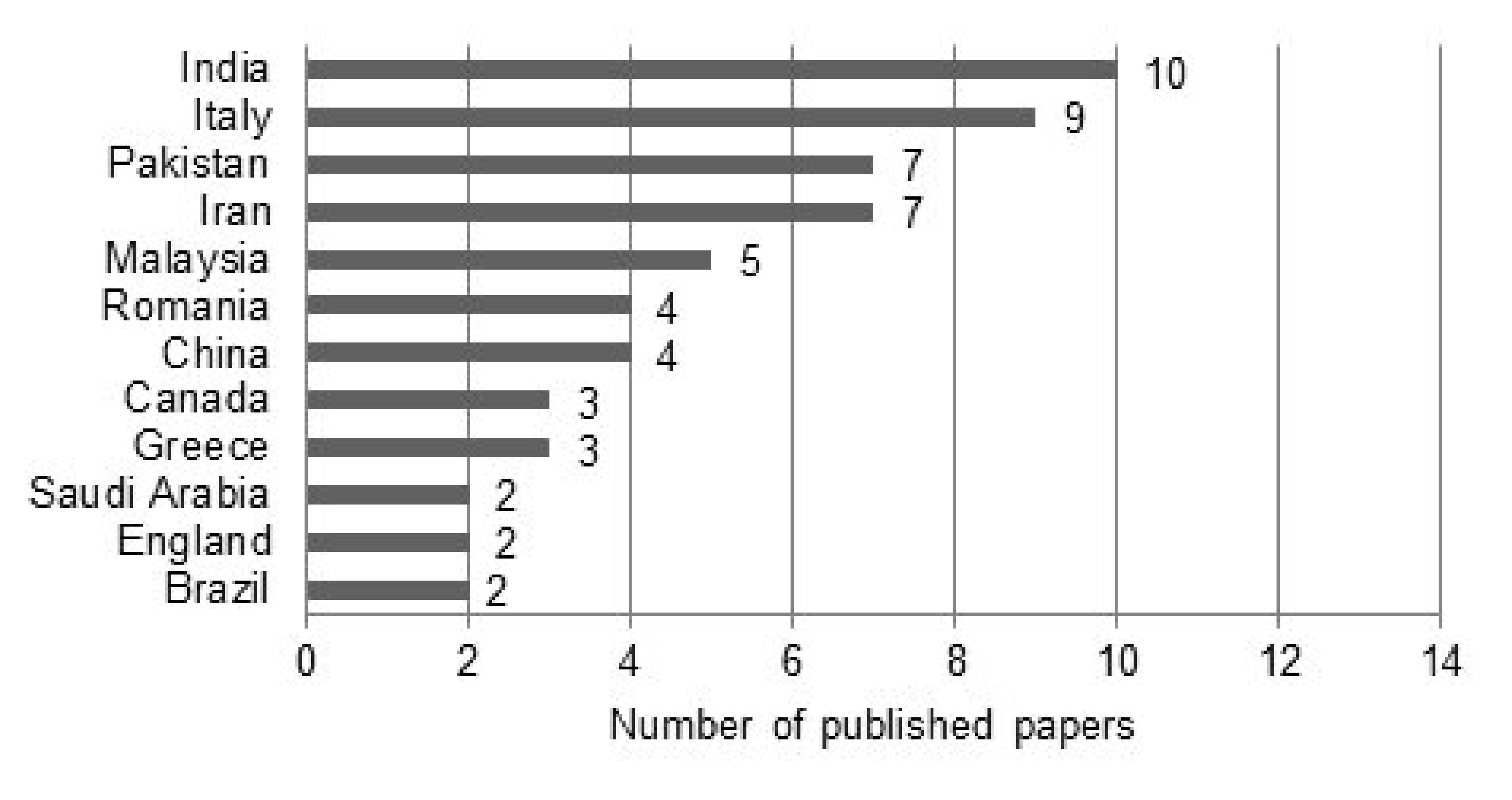}
\caption{Active countries.}
\label{Fig:8}
\end{figure}

The active countries in the published papers can also be extracted from the information presented in Table \ref{Tab:5}. Such data can be obtained based on the organizational affiliation of the authors/coauthors. Figure \ref{Fig:8} shows the countries that are most active in publishing papers on software module clustering and the share of papers published by each country with respect to the total number of publications.

The most active countries were those that published more than one research paper. Such countries produced approximately 41\% (58/143) of the total publications. India and Italy were the two most active countries, with 10 and 9 published papers, respectively. Those papers published from India were from a collaboration between the authors "Jitender Kumar Chhabra," "AmarjeetPrajapati," and "Amit Rathee."



\subsection{Applications of Software Module Clustering (RQ3)}

An in-depth analysis of the selected papers revealed that software module clustering applications can be classified into 14 areas (A1-A14). This result is presented in Table \ref{Tab:6} which shows the papers in each application area. From the identified application areas, it is clear that "information recovery (A1)" and "restructuring (A2)" are the top two applications of software module clustering with 59 and 44 published papers, respectively.


\begin{table*}
\caption{Application areas of software module clustering.}\label{Tab:6}

\centering
\begin{tabular}{ |l|p{7cm}|p{6cm}|l| }

 \hline
 ID &   Application Areas & Published Papers & Total\\
 \hline
  
  A1 &    To support information recovery such as abstraction levels (e.g., modules or design patterns as low-level abstraction and architectures as high-level abstraction). Information recovery is essential for better software understanding, maintaining, and updating. &

  \cite{ref5-2011}, \cite{ref4-2009}, \cite{ref2-2017}, \cite{ref3-2017}, \cite{ref6-2017}, \cite{ref14-2018}, \cite{ref13-2010}, \cite{ref3-2011}, \cite{ref11-2012}, \cite{ref12-2016}, \cite{ref2-2019}, \cite{}, \cite{ref13-2019}, \cite{ref14-2019}, \cite{ref1-2016}, \cite{ref6-2019}, \cite{ref11-2013}, \cite{ref9-2013}, \cite{ref10-2013}, \cite{ref3-2015}, \cite{ref1-2017}, \cite{ref3-2018}, \cite{ref4-2010}, \cite{ref4-2017},  \cite{ref4-2016}, \cite{ref6-2013}, \cite{ref12-2015}, \cite{ref9-2015}, \cite{ref3-2009},  \cite{ref6-2009}, \cite{ref5-2010}, \cite{ref7-2010}, \cite{ref8-2010}, \cite{ref9-2010}, \cite{ref11-2010},  \cite{ref7-2011}, \cite{ref10-2011}, \cite{ref11-2011}, \cite{ref13-2011}, \cite{ref3-2012}, \cite{ref6-2012}, \cite{ref7-2012}, \cite{ref7-2013}, \cite{ref8-2013},  \cite{ref12-2013}, \cite{ref13-2013}, \cite{ref5-2014}, \cite{ref2-2015},  \cite{ref5-2015},  \cite{ref3-2016}, \cite{ref5-2017}, \cite{ref12-2017}, \cite{ref14-2017}, \cite{ref8-2018},  \cite{ref17-2018}, \cite{ref18-2018}, \cite{ref11-2019}, \cite{ref12-2019}, \cite{ref7-2019}        
  &    59\\
  \hline
   A2 &    To increase software maintainability by reducing the complexity of its modules via decoupling, restructuring, or refactoring. &        
  \cite{ref11-2008}, \cite{ref5-2012},  \cite{ref4-2011}, \cite{ref7-2017}, \cite{ref10-2017}, \cite{ref10-2018}, \cite{ref3-2010}, \cite{ref9-2019}, \cite{ref4-2013}, \cite{ref13-2017}, \cite{ref6-2008},  \cite{ref6-2010}, \cite{ref11-2017}, \cite{ref12-2010},  \cite{ref12-2011},  \cite{ref9-2012}, \cite{ref7-2008}, \cite{ref10-2014}, \cite{ref14-2016}, \cite{ref13-2008}, \cite{ref1-2009}, \cite{ref5-2009}, \cite{ref9-2009},  \cite{ref2-2008}, \cite{ref5-2008},  \cite{ref8-2008},  \cite{ref12-2008}, \cite{ref10-2009},  \cite{ref9-2011}, \cite{ref10-2012}, \cite{ref8-2012}, \cite{ref1-2013},  \cite{ref6-2014}, \cite{ref8-2014}, \cite{ref9-2014},  \cite{ref12-2014}, \cite{ref7-2015}, \cite{ref11-2015}, \cite{ref11-2016},   \cite{ref9-2017},    \cite{ref4-2018},  \cite{ref3-2019}, \cite{ref4-2019}, \cite{ref8-2019}
   & 44\\ 
  \hline
  A3 &    To identify software components that implement related functionality (a component is a strongly related set of software entities such as files, classes, and modules). Software component identification is important to promote the reuse of recovered components from existing software systems as building blocks for newly developed systems. &    
  
  \cite{ref1-2011}, \cite{ref2-2014}, \cite{ref15-2019}, \cite{ref8-2011}, \cite{ref11-2014}, \cite{ref2-2016}, \cite{ref2-2012}, \cite{ref3-2013}, \cite{ref5-2013},    \cite{ref8-2009}, \cite{ref1-2015}, \cite{ref4-2015},  \cite{ref8-2016}, \cite{ref5-2018}, \cite{ref12-2018}, \cite{ref1-2019}
  
  &    16\\ 
 \hline
  
  A4 &    To identify the fault/defect proneness of software modules that require further examination and maintenance. &    
  
   \cite{ref2-2013}, \cite{ref6-2016}, \cite{ref1-2014},  \cite{ref8-2015},  \cite{ref2-2010}, \cite{ref2-2011}, \cite{ref11-2018}

  &    7\\ 
 \hline
  A5 &    To find concept locations (also called change propagation or change impact analysis) in source code. Finding concept locations in source code helps to identify where changes are to be made in response to new requirements. &    
  \cite{ref6-2011}, \cite{ref1-2012}, \cite{ref1-2018}     
  
  &   3\\ 
 \hline
  
  A6 &    
To evaluate a software system’s maintainability via determining the entities that are more difficult to maintain than others. &
  \cite{ref2-2009}, \cite{ref4-2012}, \cite{ref13-2018}
  & 3\\ 
 \hline
 
 A7 &    To map components found in the source code onto machines found in a distributed environment. 
Here, the number of clusters is created based on the number of machines in the used distributed environment.
    &     
  
  \cite{ref1-2008}, \cite{ref7-2009}
  & 2\\

 \hline
 A8 &    To identify duplicate code (also called code clone). A code clone is a piece of code that appears at least twice in a software system due to copy-paste activity or reusability of existing code. Code clone detection helps improve the performance of software and reduce its maintenance cost and effort.  &    
  \cite{ref10-2015},  \cite{ref13-2016}, \cite{ref5-2019} 
  & 3\\ 
 \hline
  A9 &    To identify groups of similar code changes. A code change is a sequence of edit operations that turns an original code (old version) into its modified version (new version). The identified code changes can serve as inputs for some software tools such as recommendation and bug fixing tools.     &  
  
  \cite{ref5-2016}
  &    1\\ 
 \hline
  A10 &    To identify different types of software architectures that are used in heterogeneous systems. &     
  \cite{ref9-2016}
  &    1\\ 
 \hline
 
A11    & To understand locations within an AOP system that could cause or exhibit aspect interference problems. Thus, it is essential to develop an interference-free AOP system. &     

\cite{ref8-2017}
&    1\\
\hline

A12    & To discover program topoi (which are summaries of the main capabilities of a program).     &     
\cite{ref7-2018}

& 1\\
\hline
A13    & To generate microservice candidates from a monolithic software system.    &     

\cite{ref9-2018}
& 1\\
\hline
A14    & To assess the initial partition of a software system (assess the extent to which a certain partition allows its parts to evolve independently).     &     

\cite{ref4-2008}
& 1\\
\hline
\end{tabular}
\end{table*}


\subsection{Software Module Clustering Process (RQ4)}

\begin{figure}
\centering
\includegraphics[scale=0.65]                {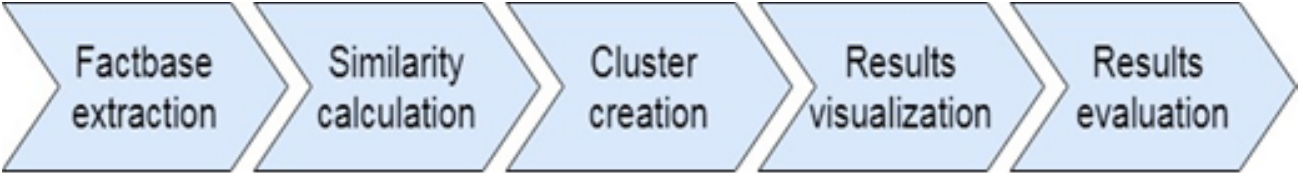}
\caption{Software module clustering process.}
\label{Fig:9}
\end{figure}

Answering RQ4 helps to determine the detailed steps of the software module clustering process. Moreover, it helps to identify the algorithms, techniques, and tools used in each step of the process. The ideal software module clustering process includes five main steps, as shown in Figure \ref{Fig:9}. 
Here, the process ends when reaching a stopping criterion such as a maximum number of iterations or a desired number of clusters. The following subsections describe the steps of the process.

\subsubsection{Factbase Extraction}

The input that a software clustering algorithm expects is often called a factbase as it contains facts (e.g., relationships between software entities) extracted from the target software system \cite{other8}. In this step, the extracted factbase should consist of sufficient information about the target software system to ensure meaningful clustering \cite{ref7-2008}. Factbase extraction includes target software system selection, factbase source selection, filtering and preprocessing, entity selection, and feature selection.
\\

\noindent\textbf{A. Target software system selection (RQ5)}

\begin{figure}
\centering
\includegraphics[width=3.4 in]                {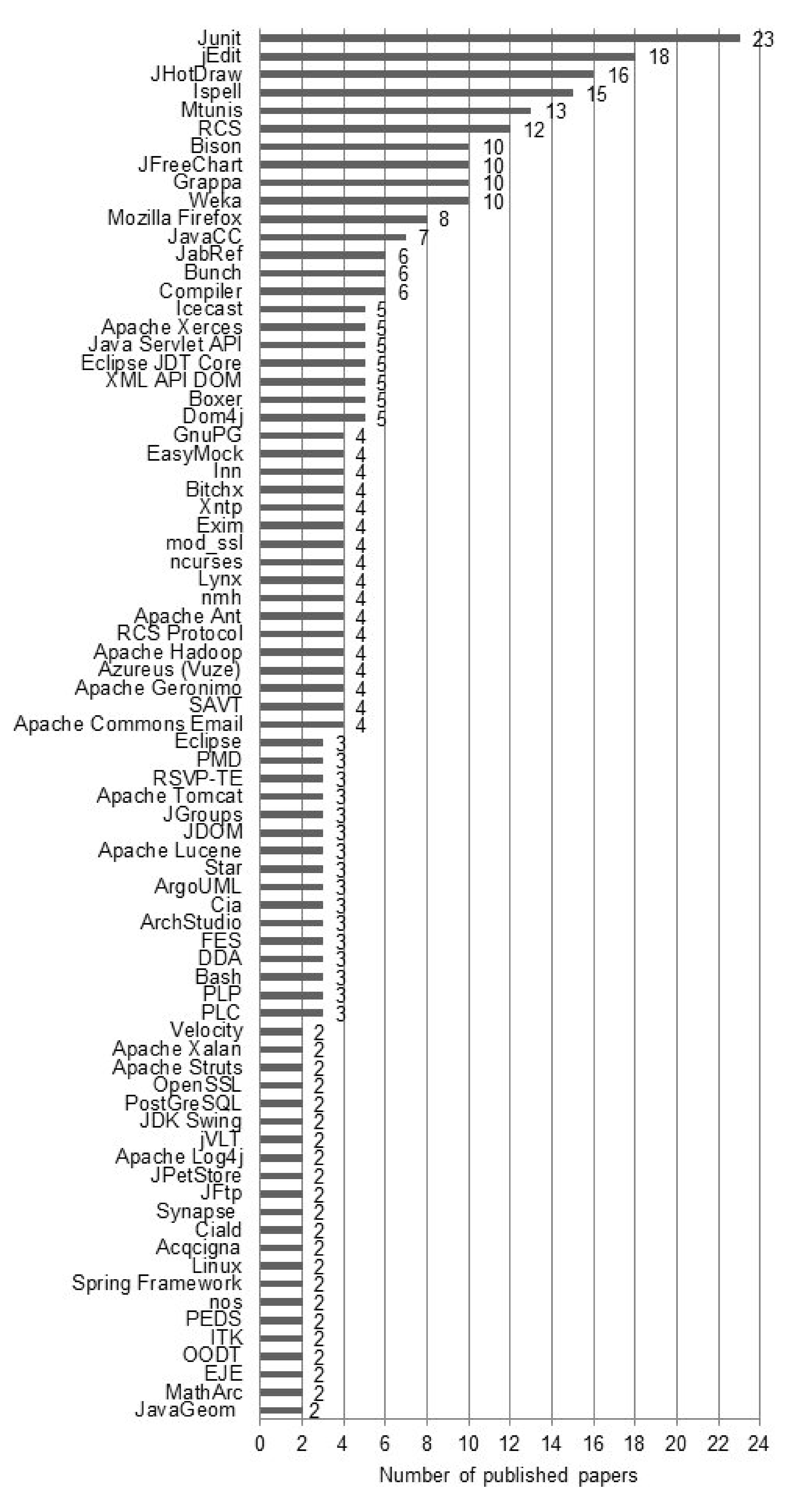}
\caption{Number of published papers vs. the targeted software systems.}
\label{Fig:10}
\end{figure}

Before starting the software clustering process, target software systems must be specified. The analysis of the selected studies reveals that open-source software systems written in Java and C/C++ have been the focus in the literature. Figure \ref{Fig:10} shows the most commonly used target software systems.

The figure clearly shows that "Junit," "jEdit," and "JHotDraw" are the most commonly used systems in the experimental tests, accounting for approximately 16\% (23/143), 13\% (18/143), and 11\% (16/143) of the published papers, respectively. In addition, the top three systems are written in Java. The reason for this fact could be that Java is still the most widely used programming language according to the TIOBE Index \cite{tiobe}. For the experimental clustering tests, we recommend selecting the target systems that are well-known to the developer and researcher communities. They are free and open-source, written in widely used programming languages, and updated frequently. \\

\noindent\textbf{B. Factbase source selection (RQ6)}

Any software clustering process starts with the construction of a factbase. As mentioned, the factbase contains information on the target software system, such as software entities (e.g., classes and variables) and their relationships (e.g., inheritance and method calls). Such information can be extracted from various sources.
Table \ref{Tab:7} presents in detail the different types of factbase sources. After constructing the factbase, a software clustering algorithm can be applied to group entities from the factbase into useful subsystems. Many clustering methods combine facts extracted from different sources to obtain reasonable results at the cost of more complex data processing \cite{ref7-2010}.

\begin{figure}
\centering
\includegraphics[width=3.4 in]                {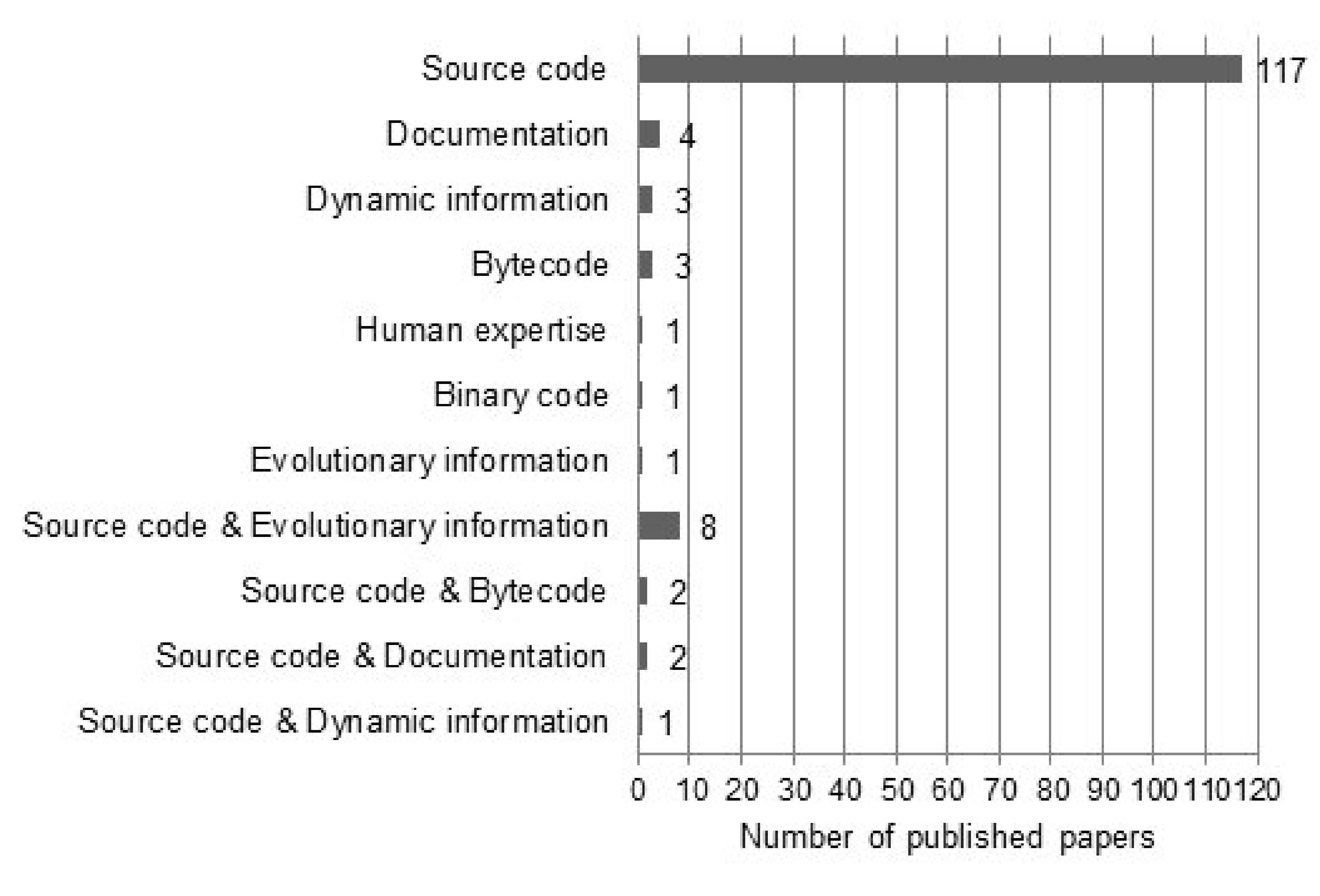}
\caption{Number of published papers vs. the factbase source.}
\label{Fig:11}
\end{figure}

\begin{figure}
\centering
\includegraphics[width=3.4 in]                {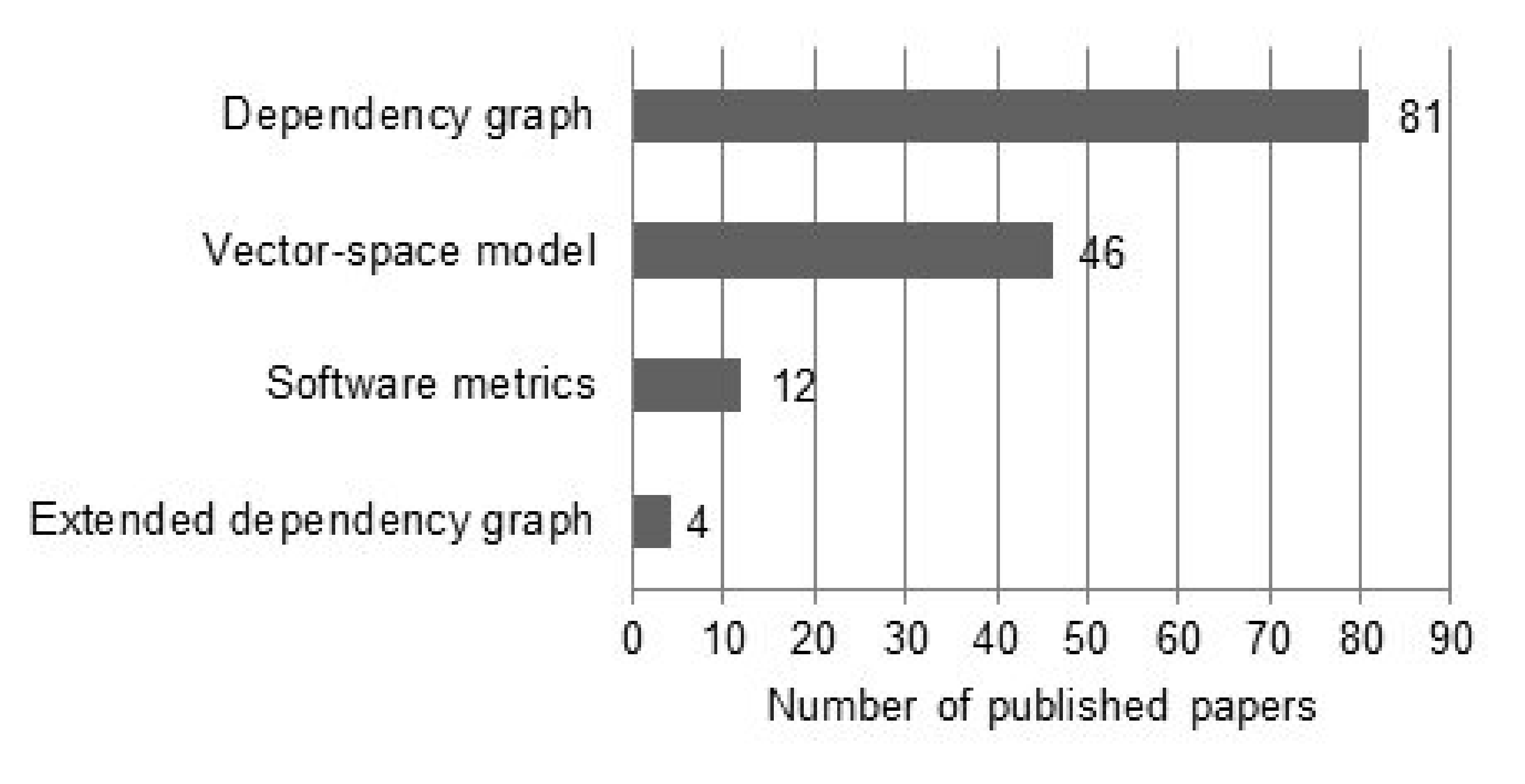}
\caption{Number of published papers vs. the factbase extracted.}
\label{Fig:12}
\end{figure}

\begin{landscape}
\begin{table}
\caption{Factbase sources.}
\centering 
\scriptsize
\begin{tabular}{ |p{2cm}|p{9cm}|l|p{10cm}| }
 
\label{Tab:7} 
\\
\hline
 Factbase Sources &   Factbase Extracted & Factbase Type & Major Drawbacks\\
 \hline
  Source code &    •    Structural dependencies (also called syntactic information) between software entities (e.g., function calls, class references, class inheritance, file inclusions, package usages, and global variable access). Often represented as directed graphs or matrices.
•    Domain or naming information (also called lexical information or semantic information) found in comments, variable names, object names, method names, class names, file names, and source code statements    & Static    & •    It is language-dependent.
•    May be mismatched to a product version of binary modules because of errors, and updates
•    Comments and naming conventions may not be used or followed by software engineers.
•    Extracting domain knowledge is known to be a difficult task due to the existence of noise in the data. 
•    It deals with structural relationships rather than behavioral relationships (dynamic interactions among system entities).
•    Not available with some software systems.
•    If the software system is extremely large. The number of entities involved in clustering would be so large that it could be unmanageable to track.
•    Natural language processing techniques (stemming) to extract naming information is very time-consuming
\\

 \hline
  Binary code (executable files)    & •    Compilation parameters, linkage parameters, and symbol table that contain relationships between components and modules of a software system.    &     Static    & •    It is language-dependent.
•    It is a compiler dependent. Thus, compiling a source code via two different compilers may result in two different binary codes. This is because each compiler has its building parameters and nature of work.
\\
  \hline
  Bytecode    & •    Class dependencies.
•    Names of variables and classes    &    Static    &It is language-dependent.
•    Not available with some software systems. 
 \\ 
  \hline
  Documentation    &•    System's structure or design.
•    System's functional and non-functional requirements.    &    Static    & •    Facts are often incomplete, outdated, unsynchronized, vague, or even missing due to long term maintenance or personnel turnover. Thus, they may not reflect the current state of the system. Also, documentation could be completely missing.
\\ 
 \hline
 Human expertise    & •    Knowledge of requirement documents.
•    High-level design.
•    Knowledge of architectural styles.    &    Static    &•    On the same software system, different experts may construct different software decomposition.
 \\ 
 \hline
  Configuration files    &•    Declaration of used resources.
•    Security permissions.    &    Static    &•    Unavailable because they are stored in middleware configuration files that are not part of the the software systems' source or binary code.
\\ 
 \hline
  Dynamic information    & •    Object construction and destruction. 
•    Sources of exceptions and errors. 
•    Method entry and exit. 
•    Method invocations/calls. 
•    Execution logs (structured execution outputs that capture the process of building, deploying and testing various scenarios) 
•    Performance counters and statistics such as number of threads, size of buffers, number of network connections, CPU and memory usage, number of component instances, average, maximum, and minimum response time.    &    Dynamic    &•    It is language-dependent.
•    Often, the source code does not remain untouched.
\\ 
 \hline
  Data files    &•    Such as database, text, and reports files used as input or output that contain information about which classes or modules perform file operations (e.g., create, read, and write).    &    Dynamic    &•    Some software systems do not deal with input and output files.\\ 
 \hline
 Files organization    & •    The organization of software classes, files, folders, packages and their physical locations may provide information about their responsibilities.    &    Organizational    &•    Often, software files and folders are not organized well. \\ 
 \hline
  Human organization    &•    The structure of the system. Usually, a developer is responsible for associated components.    &    Organizational    &•    In medium and large software systems, it is not identified well due to developers turn out. \\ 
 \hline
 Evolutionary (historical) information    &•    File ownership, file timestamps, file versions, commit logs, persons making changes, maintenance activities, and bug report information extracted from software repositories, version control/management systems, bug/issue tracking systems, or release documents.
•    Also, it tells us how, when, and why a given software entity was modified.    &    Evolutionary    &•    It does not cover the essential parts of the software system unless huge evolutionary data has to be dealt with. Thus, it is usually difficult to deal with due to the issue of size.
•    The lack of evolutionary data formatting.
\\
\hline
\end{tabular}
\end{table}
\end{landscape}


Figure \ref{Fig:11} shows that "source code", "documentation", "dynamic information", and  "bytecode" are the most commonly used sources for factbase extraction, constituting approximately 82\% (117/143), 3\% (4/143), 2\% (3/143), and 3\% (2/143) of the total published papers, respectively. In addition, many papers combined two sources for factbase extraction. In this respect, "source code and evolutionary information" is the most commonly used combination, accounting for approximately 6\% (8/143) of the total publications.

The factbases are extracted from the sources in different forms. Analysis of the selected papers reveals that the "dependency graph," "vector-space model," "software metrics," and "extended dependency graph" are the most commonly used forms of the factbase. Figure \ref{Fig:12} shows the analysis results of the published papers. The figure also shows that approximately 57\% (81/143), 32\% (46/143), 8\% (12/143), and 3\% (4/143) of the total published papers considered "dependency graph", "vector-space model", "software metrics", and "extended dependency graph", respectively. The following points describe these factbase forms in detail:

\begin{itemize}
\item \textbf{Dependency graph:} It is a graph representation of the target software system. The nodes in the dependency graph represent software entities, whereas the edges represent the logical/static relationships between entities. In some cases, edges are weighted to denote the degree of dependency. Once the dependency graph is extracted, many characteristics of the target software system can be discovered, such as the independence degree of the software entities based on their relationships.

\item \textbf{Vector-space model:} It is used to capture the relative importance of terms (e.g., class name, function name, object name, and variable name) in a document, e.g., class file, and program file. In the vector-space model, a document is represented by a vector of terms extracted from the document with associated weights (which can often be computed using the term frequency-inverse document frequency (TF-IDF) method \cite{ref6-2011}) representing the importance of the terms in the document and within the whole document collection (the target software system).

\item \textbf{Software metrics:} They are quantitative measures that enable software engineers and managers to understand the target software system. The number of code lines per class, number of methods per class, and depth of inheritance level are possible metrics.

\item \textbf{Extended dependency graph:} It is a graph that combines logical/static relationships and evolutionary relationships among software entities. The evolutionary relationships of the targeted software system represent the changes applied to its source files over time. Currently, many version control systems, such as the concurrent versions system (CVS) and Git, store these changes.

\end{itemize}

Notably, researchers typically use tools, which are mostly open-source Java programs, to perform factbase extraction. Table \ref{Tab:8} presents all the tools that have been used in the selected studies, along with their links.\\

\begin{table*}
    \centering
    \scriptsize    

\captionof{table}{Factbase Extraction Tools.} 
\label{Tab:8}
\begin{tabular}{ |l|p{11cm}| } 
 \hline
Tool & URL\\
\hline 
Dependency Finder & http://depfind.sourceforge.net/\\
\hline 
Class Dependency Analyzer (CDA) & http://www.dependency-analyzer.org/\\
\hline 
ASM & http://asm.ow2.org\\
\hline 
CScout & https://www.spinellis.gr/cscout/\\
\hline 
Bunch & https://wiki.eecs.yorku.ca/project/cluster/tools\\
\hline 
Sotograph & https://www.hello2morrow.com/products/sotograph\\
\hline 
Structure101 & https://structure101.com/\\
\hline 
Understand 2.0 & https://scitools.com/\\
\hline 
JRipples & https://marketplace.eclipse.org/content/jripples\\
\hline 
E-Quality & http://smart.cs.itu.edu.tr/tools/equality/\\
\hline 
Javassist & http://www.javassist.org/\\
\hline 
Visual Paradigm & https://www.visual-paradigm.com/\\
\hline 
Doxygen & http://www.doxygen.nl/\\
\hline 
Jdeps & https://docs.oracle.com/javase/8/docs/technotes/tools/unix/jdeps.html\\
\hline 
Roslyn & https://github.com/dotnet/roslyn\\
\hline 
Stan4J & http://www.stan4j.com/\\
\hline 
PomWalker & https://github.com/raux/PomWalker\\
\hline 
SrcML.NET & https://github.com/abb-iss/SrcML.NET\\
\hline 
\end{tabular}
\end{table*}


\noindent\textbf{C.    Filtering and preprocessing}

Filtering is a useful preprocessing phase in any clustering process to identify and remove unnecessary textual and nontextual information that has been extracted from comments and source codes. Textual information can be meaningless words, such as words with less than three characters, language keywords, or common English words that are not usually useful for a search \cite {ref3-2011}.
Textual information can also be library classes or header files used in multiple modules and made available across the implementations of a programming language. These classes have to be eliminated; otherwise, they tend to group many classes in a single large cluster around them \cite{ref13-2010}. Nontextual information can be operators, symbols, special characters, and punctuations. Preprocessing can be implemented in the form of a normalization procedure. The attributes of the software entities can be a mixture of numerical and categorical data. Here, with the help of normalization, all the attributes are treated equally \cite{ref3-2016}. Preprocessing can also be conducted in the form of dimensional reduction, which is used to reduce the size of the input vectors \cite{ref11-2018}. Filtering is an essential phase to facilitate further processing and avoid the risk of decreasing the clustering quality \cite{ref7-2010}.\\


\noindent\textbf{D.    Entity selection}

Software entities are the input to any software clustering technique. Therefore, entities to be clustered must be identified beforehand. The selection of entities depends mainly on the aim of the clustering technique \cite{ref12-2010}. Table \ref{Tab:9} presents two examples in this respect.

\begin{table*}
    \centering

\captionof{table}{Entities selection.} 
\label{Tab:9}
\begin{tabular}{ |l|p{6.5cm}|p{6cm}| } 
 \hline
 Clustering aim & Input entities & Entities abstraction level  \\
 \hline
 Software comprehension &   Functions and their call statements & Low-level (also called detailed level)\\ 
 \hline
 Architecture recovery &  Software classes, packages, modules, and files & High-level (also called architectural level)\\ 
 \hline
\end{tabular}
\end{table*}

Low-level entities represent the functionality of the target software system much more clearly than high-level entities. However, some software systems are large and contain enormous numbers of functions that make the use of functions in the clustering process inappropriate. In such cases, the clustering of high-level entities is preferable.\\

\noindent\textbf{E.    Feature selection}

Each software entity has a set of features. The features of a class entity, for example, can be broadly divided into two types \cite{ref1-2011, ref7-2013}:

\begin{itemize}
\item Nonformal features: These include file naming convention, class creation date, number of functions, number of variables, number of lines of codes, and comments. Nonformal features do not directly affect system behaviors. Also, they can be easily extracted, interpreted, and understood by humans.

\item Formal features: These include class invocations, method invocations, and entity relationships. Relationships can be categorized into two types \cite{ref12-2010}:

\begin{itemize}
\item Direct relationships: Represent an immediate connection between two entities e.g., if function $f_1$ calls another function $f_2$, then $f_1$ and $f_2$ are directly related.
\item Indirect relationships: Represent the proportion of common features that two entities share e.g., if functions $f_1$ and $f_2$ both call function $f_3$, then $f_1$ and $f_2$ are indirectly related to each other. 
\end{itemize}

The formal features have a direct impact on system behaviors. For example, if a class changes its invocation from one class to another, then changes in the system's behavior should be expected. Extracting formal features, however, can be more complicated, as parsing rules need to be applied. This process becomes even more complicated when only partial information is available.

\end{itemize}


Feature selection aims to prepare the features of all software entities for the next step as a feature matrix (also called the entity-feature matrix) \cite{ref8-2010}. A feature matrix is a two-dimensional matrix where the rows represent software entities and the columns represent their features. The value of each cell in the matrix is either 0, which indicates the absence of a feature, or 1, which indicates the presence of a feature \cite{ref8-2009}. Table \ref{Tab:FM} presents a feature matrix of software containing 5 entities and 4 features. Feature F1 is present in all the entities, while feature F3 is absent in all the entities. Features F2 and F4 are present or absent in the entities.

\begin{table}[h]
    \centering
\captionof{table}{A simple feature matrix.} \label{Tab:FM}
\begin{tabular}{ |l|l|l|l|l| } 
 \hline
  & F1 & F2 & F3 & F4 \\
 \hline
  E1 & 1 & 0 & 0 & 1\\ 
 \hline
  E2 & 1 & 0 & 0 & 0\\ 
  \hline
  E3 & 1 & 1 & 0 & 1\\
  \hline
  E4 & 1 & 1 & 0 & 0\\
  \hline
  E5 & 1 & 1 & 0 & 1\\ 
 \hline
\end{tabular}
\end{table}

Some clustering approaches apply weighting schemes (e.g., binary weighting, absolute weighting, and relative weighting) to the features to represent the significance of each one. Thus, the connection strength between a pair of entities can be calculated \cite{ref1-2011}.

\subsubsection{Similarity Calculation (RQ7)}

\begin{figure}
\centering
\includegraphics[width=3.4 in, height=6cm]                {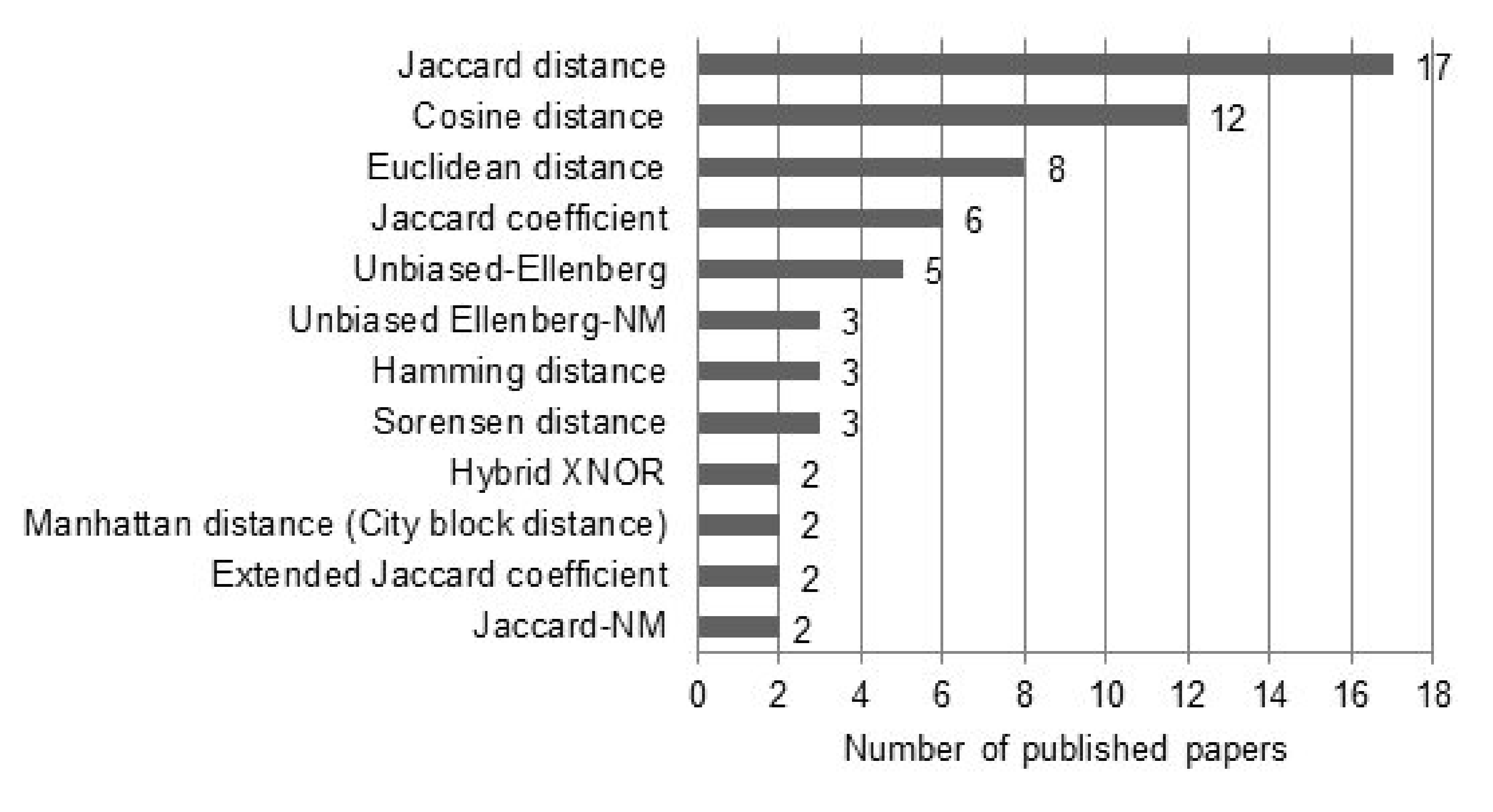}
\caption{Number of published papers vs. similarity/dissimilarity measure.}
\label{Fig:13}
\end{figure}

Similarity measures are used with software module clustering to determine the most similar or dissimilar entities based on their features. Generally, entities are considered more similar if they share more common features \cite{ref13-2010}. After the computation of the similarity among all pairs of software entities, a similarity matrix can be generated for the next step. The most commonly used measures in the considered papers are shown in Figure \ref{Fig:13}. Clearly, "Jaccard distance," "Cosine distance," and "Euclidean distance" are the most common, accounting for approximately 12\% (17/143), 8\% (12/143), and 6\% (8/143) of the total publications, respectively.

\subsubsection{Cluster Creation (RQ8)}

Here, a clustering algorithm must be applied to similar group entities of the target system based on specific features.
The selection of a suitable algorithm to apply to a task is difficult. However, the authors in \cite{other8} introduced a method to help in this respect.

The analysis of the considered published papers in the study reveals the following general categories of clustering algorithms used in the literature:

\begin{itemize}
    \item  \textbf{Hard clustering: }In this type, a software entity can be in only one cluster and can be divided into two types:

\begin{itemize}
\item Hierarchical clustering  \cite{ref11-2008}: For a given set of $n$ entities, hierarchical clustering algorithms are divided into two types: 

\begin{itemize}

\item Agglomerative (bottom-up) algorithms: The algorithms begin with $n$ singletons (each of one entity) and merge them until a single cluster is obtained. The two most similar clusters are merged in each step of the process. Agglomerative algorithms provide different perspectives of software clustering: the earlier clustering iterations present a detailed view of the software architecture and the later ones present a high-level view \cite{ref1-2011}.    

\item Divisive (top-down) algorithms: These algorithms start with one cluster (containing all $n$ entities) and split it until $n$ clusters are obtained.
\end{itemize}

\item Partitional clustering: For a given set of n entities, this approach simply divides the set of entities into nonoverlapping clusters such that each entity is in exactly one cluster. 
\end{itemize}

Notably, some studies have combined more than one clustering algorithm. For example, hierarchical and partitional clustering can be combined to achieve a common goal. This type of clustering is called cooperative clustering \cite{ref9-2010}.
\newline

\item \textbf{Soft clustering} (also called fuzzy clustering): This approach is based on fuzzy logic, where a software entity can be in one or more clusters. Thus, a probability or a membership degree of that entity in those clusters is assigned \cite{ref3-2011}.

\end{itemize}

\begin{figure}[h]
\centering
\includegraphics[width=3.4 in, height=6cm]                {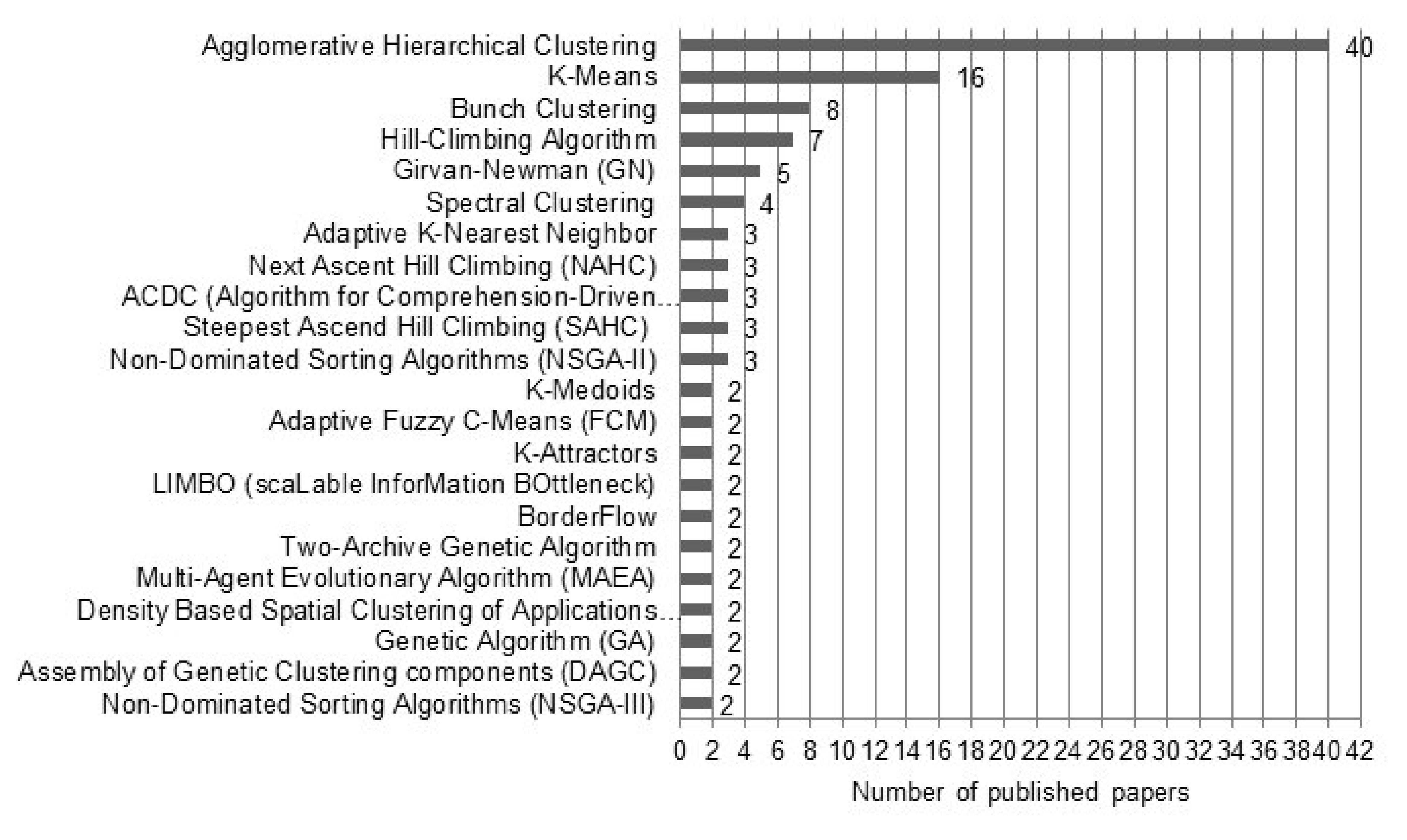}
\caption{Number of published papers vs. clustering algorithm.}
\label{Fig:14}
\end{figure}

A deep analysis of the selected papers reveals that twenty-two clustering algorithms are the most commonly used, as shown in Figure \ref{Fig:14}. Notably, "agglomerative hierarchical clustering" and "k-means" are the most commonly used clustering algorithms in the context of software module clustering, with percentages of approximately 28\% (40/143) and 11\% (16/143) of the total publications, respectively. Figure \ref{Fig:14} clearly shows this result. 

\begin{figure}[h]
\centering
\includegraphics[width=3.4 in]                {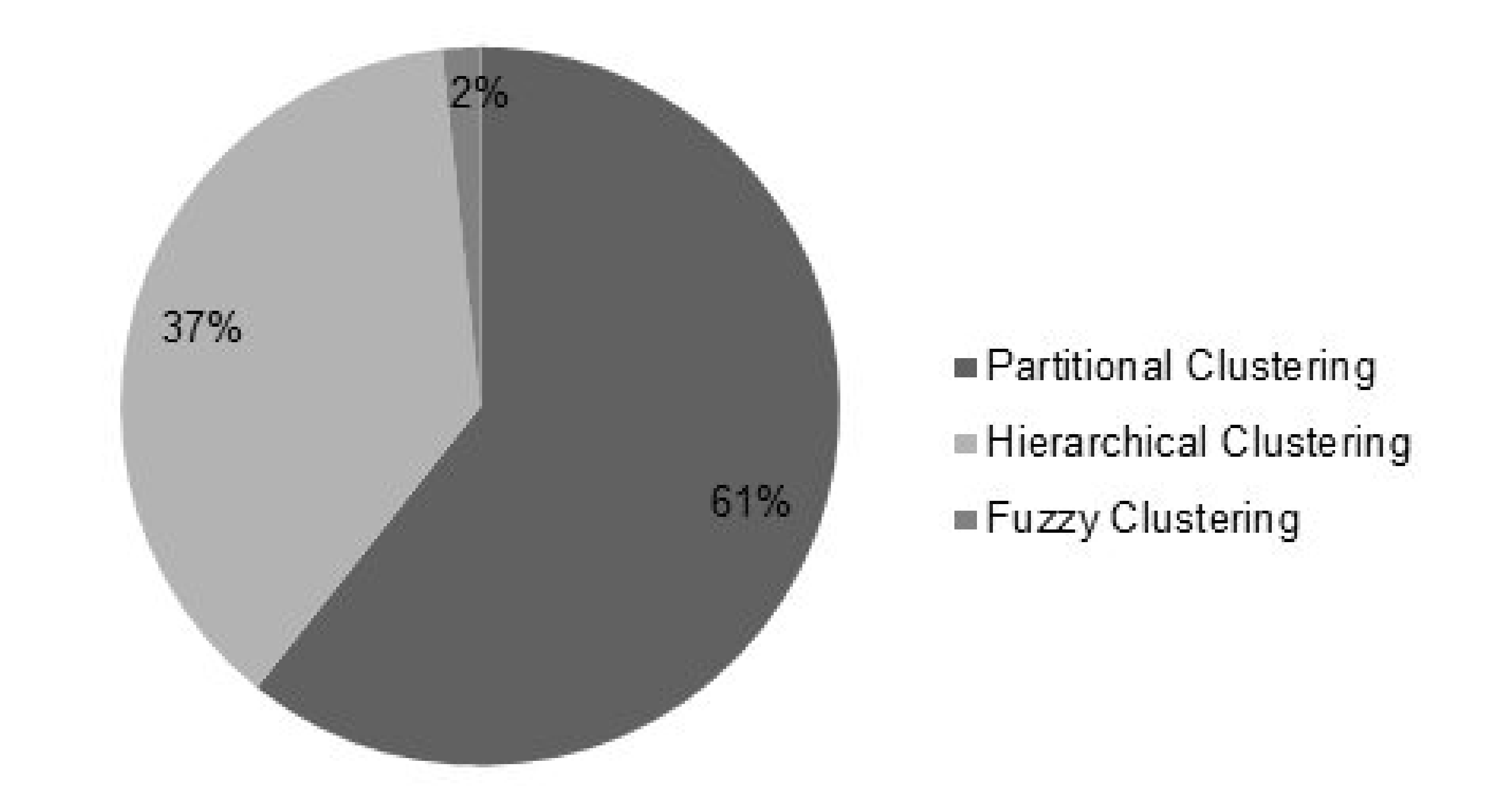}
\caption{Number of published papers vs. clustering type.}
\label{Fig:15}
\end{figure}

Figure \ref{Fig:15} shows that "Partitional Clustering" is the most commonly used type of clustering, accounting for approximately  50\% (71/143) followed by "Hierarchical Clustering", with approximately 31\% (44/143), and "Fuzzy Clustering", with approximately 1\% (2/143) of the total publications.

\begin{figure}[h]
\centering
\includegraphics[width=3.4 in]                {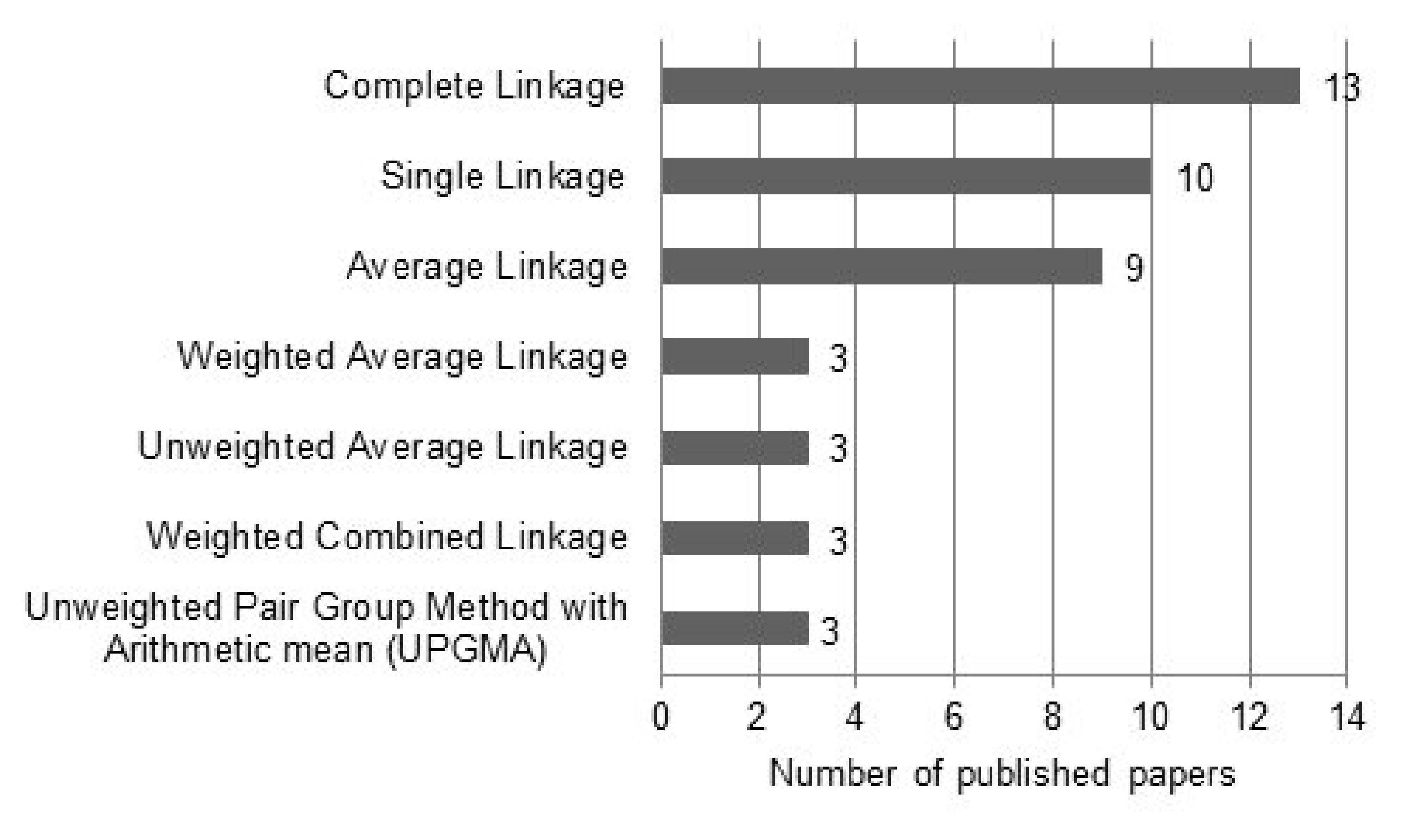}
\caption{Number of published papers vs. distance computation methods in hierarchical clustering.}
\label{Fig:16}
\end{figure}

In hierarchical clustering, several methods can be used to compute the distance between clusters. An examination of the selected studies reveals the seven most commonly used methods. Figure \ref{Fig:16} shows this result. Clearly, the top three methods are "complete linkage", with approximately 9\% (13/143), "single linkage", with approximately 7\% (10/143), and "average linkage", with approximately 6\% (9/143) of the total publications.

The analysis of the selected studies indicated the use of several different termination conditions. The clustering process can be terminated based on one of the following cutoff conditions \cite{ref8-2008, ref1-2011}:

\begin{itemize}
\item
A good solution has been reached 
(e.g., when satisfying results are obtained by applying one or more clustering evaluation metrics).
\item
A maximum number of iterations has been reached. 
\item
Improvement has not been found for a long time. 
\item
All software entities are combined into a single cluster.
\item
A pre-specified termination condition has been met 
(e.g., a specific number of clusters has been formed).
\end{itemize}



\subsubsection{Results Visualization of the Clustering (RQ9)}

Visualization is used after the software module clustering process to view the clustering results as graphs, dendrograms, or distribution maps. Visualization of the results is also used to enable software engineers to efficiently and conveniently examine the clustering results. To achieve this goal, the visualization tools used in the literature ease and automate the visualization process. For example, graph visualization tools compile a graph description language and generate an image file outlining the subsystems as the output of software clustering \cite{ref5-2011}. Analysis of the selected studies reveals that researchers have not focused on the visualization of the clustering results. A potential reason for this situation is related to the software module clustering systems' end-users, which are the developers themselves or domain experts. However, Table \ref{Tab:10} shows the visualization tools used in three of the studies.

\begin{table*}
    \centering
    \captionof{table}{Results Visualization Tools.} 
\label{Tab:10}
\begin{tabular}{ |l|l|l| } 
 \hline
  Tool & URL & Supported Languages \\
 \hline
  Prefuse & http://vis.stanford.edu/papers/prefuse   & Java\\ 
 \hline
  Sotograph &  http://www.hello2morrow.com/products/sotograph &  C/C++, C\#, Java, PHP, and Typescript \\ 
 \hline
  Graphviz &  http://www.graphviz.org & C/C++, Java, PHP, Python, Ruby,  Perl, Guile, and TCL\\ 
 \hline
\end{tabular}
\end{table*}

\subsubsection{Results Evaluation Metrics of the Clustering (RQ10)}



The software clustering results should be evaluated in accordance with specific criteria to assess their quality \cite{ref5-2011}. Several methods in the literature can be used to assess the quality of software clustering algorithms. Although crucial, validation of the results produced by these algorithms is difficult. Figure \ref{Fig:17} shows the most commonly used evaluation methods in the considered papers. The following points summarize these evaluation methods:

\begin{figure}[h]
\centering
\includegraphics[width=3.4 in]                {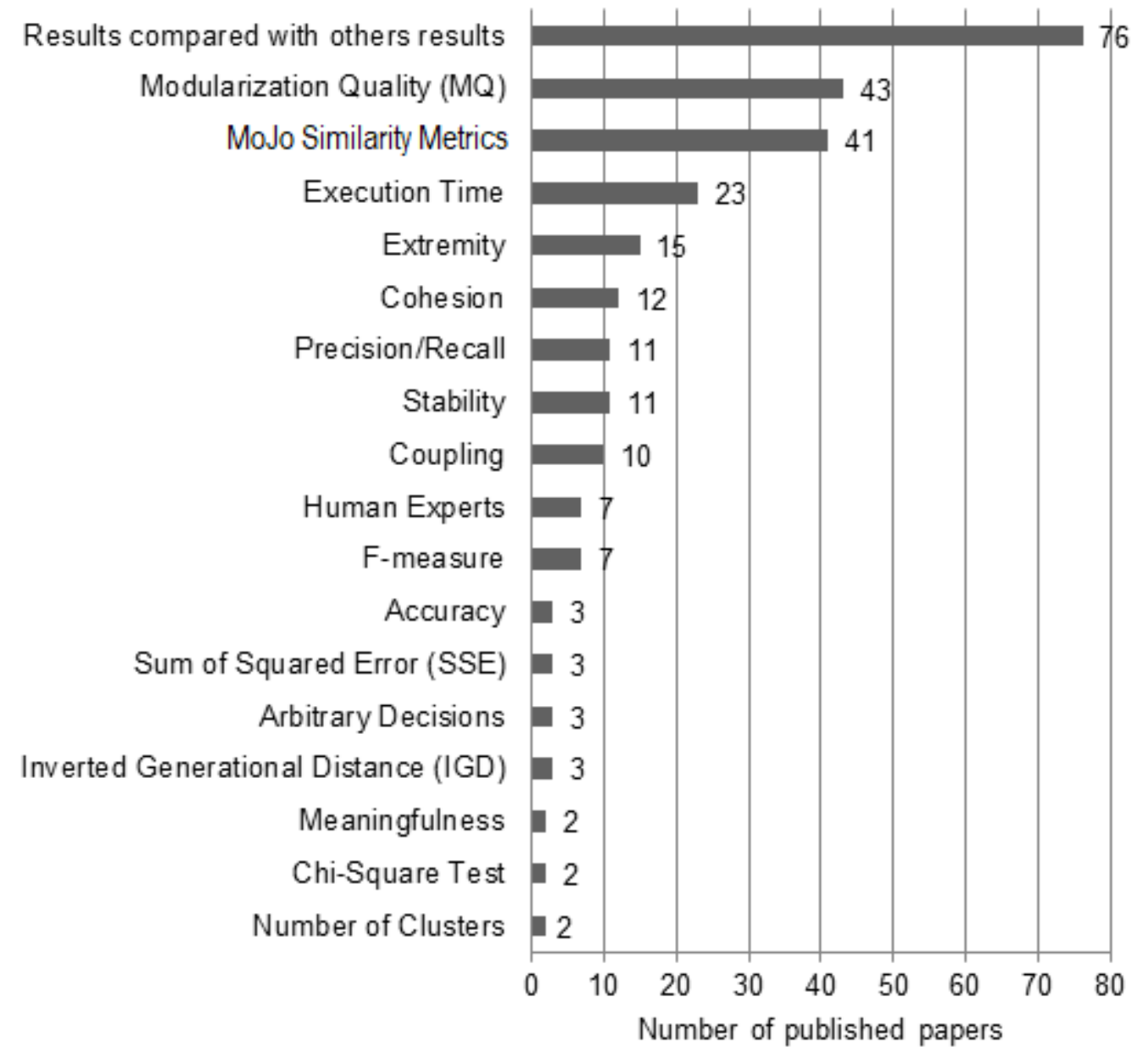}
\caption{Number of published papers vs. clustering evaluation metric.}
\label{Fig:17}
\end{figure}


\begin{itemize}
\item
Results Compared with Other Results: The results obtained from the applied clustering algorithm are compared to those of previously published studies.

\item Modularization Quality (MQ): MQ measures the cohesion and coupling of modules and is used to ensure that the clusters of a system are the cohesive modules in the clusters and the loose connections between clusters.

\item MoJo Similarity Metrics:  They measure the similarity between the partition produced by the software clustering process and the partition created by an expert (also called the expert decomposition, authoritative decomposition, benchmark decomposition, reference decomposition, gold decomposition, ground truth decomposition, or baseline decomposition) using MoJo metric and its family based on the number of Move and Join operations required to transform one decomposition into the other \cite{ref5-2011}. The similarity between the two partitions should be as high as possible.

\item Execution Time: The clustering process should not require a very long time \cite{ref5-2011}.

\item Extremity: Also called Nonextremity cluster distribution (NED), it measures whether the clusters of the partition produced by the software clustering process have extreme values or not. A good clustering process should not produce a partition that includes 
(a) Enormous clusters – clusters with too many software entities would reduce the cohesion.
(b) Singleton clusters (also called isolated clusters) or small clusters – clusters with one or too few software entities would increase the coupling of software \cite{ref11-2016}.

\item Cohesion (intraconnectivity): It measures the density of connections among the software entities in a single cluster. High cohesion indicates good clustering, as highly dependent modules are grouped into the same cluster \cite{ref8-2016}. 

\item
Precision/Recall: Precision measures the percentage of entity pairs (in the same cluster) produced by the clustering algorithm that are also in different clusters in the expert decomposition. High precision indicates good clustering. Recall measures the percentage of entity pairs (in different clusters) in the expert decomposition that were also found by the clustering algorithm. High recall indicates good clustering \cite{ref8-2008}.

\item
Stability: It measures whether the clustering process produces similar partitions in the case of small changes between successive versions of an evolving software system \cite{ref5-2010}. 

\item
Coupling (interconnectivity): It measures the density of connections among software entities in different clusters. Low coupling indicates good clustering, as the clusters are highly independent of each other \cite{ref8-2016}.

\item
Human Experts: One or more domain experts manually evaluate the results obtained from the applied clustering algorithm. 

\item
F-Measure: It measures the goodness or accuracy of the clustering methods by calculating the weighted average of recall and precision \cite{ref8-2015}. 

\item
Accuracy: The percentage of software entities that are correctly classified.

\item
Sum of Squared Error (SSE): The sum of the squared differences between entities in each cluster generated by a clustering algorithm and the entities of each cluster in the expert decomposition. Thus, SSE can be used as a measure of variation between clusters. For example, if the SSE between entities of two clusters is equal to 0, then they are matched perfectly and have no error. Smaller SSE is better, and obtaining clusters that minimize the SSE is always desirable \cite{ref4-2015}. 

\item
Arbitrary Decisions: An arbitrary decision is required when two or more entities hold the same similarity. Thus, the percentage of arbitrary decisions is employed to evaluate the clustering \cite{ref8-2010}. 

\item Inverted Generational Distance (IGD): It is calculated as the average Euclidean distance from each reference point (true Pareto front) to the nearest solution (Pareto front obtained by the algorithm) in the solution set. Here, the set of Pareto optimal solutions produced by all algorithms overall runs is used as the true Pareto front \cite{ref3-2017}.

\item
Meaningfulness: Generated clusters should resemble the subsystems of the original system \cite{ref5-2011}.

\item
Chi-Square Test: It determines if the entities of two different clusters are related in terms of some features.

\item 
Number of Clusters: It measures the compactness of the clusters created during the software clustering process. A high number of clusters indicates that they are highly compact (cohesive), while a low number of clusters indicates that they are noncohesive \cite{ref13-2017}.

\end{itemize}


Figure \ref{Fig:17} shows that "Results Compared with Others Results" is the most commonly used clustering evaluation method, accounting for approximately 53\% (76/143) of all publications, followed by "Modularization Quality", 30\% (43/143), and "MoJo Similarity Metrics", 16\% (23/143). Many studies use more than one clustering evaluation method; thus, the results overlap.

Some clustering evaluation methods (e.g., MoJo Similarity Metrics and precision/recall) work only when expert decomposition is available as a comparison standard. Examination of the total publications reveals that a number of approaches have been presented for obtaining expert decomposition. A summary of these approaches is presented as follows:

\begin{itemize}
\item
Domain Expert-Based Decomposition: The decomposition is performed by software domain experts. Domain experts are personnel with experience in software design and development \cite{ref13-2017}.
The experts either evaluate the results produced by a clustering algorithm (e.g., if the results have a positive impact on the system’s understandability) or they provide a clustering benchmark that can be compared with the results produced by a clustering algorithm \cite{ref13-2010}. Well-known drawbacks of this approach are as follows: (a) Experts may provide many valid ways to decompose a software system into meaningful subsystems \cite{ref9-2010}. (b) They might lead to poor decomposition if they did not fully understand the purpose of the clustering approach. (c) They may lead to poor decomposition if their experience and knowledge are not sufficient \cite{ref2-2008}. In addition, finding domain experts with suitable experience, especially on open-source software systems \cite{ref3-2010} and legacy systems \cite{ref7-2012}, is difficult. (d) Software systems are constantly evolving and maintaining an up-to-date expert decomposition can be a tedious, error-prone, and time-consuming task \cite{ref7-2013}.
\item
Factual Information-Based Decomposition: The decomposition is obtained using the current factual information of the targeted software system, e.g., the folder structure, the package structure, or the file structure. 

The advantage of using this approach is that its decompositions have good quality because they have been created by the original developers and domain experts \cite{ref5-2010}.
\item
Original Developer-Based Decomposition: The decomposition is obtained by approaching the original developers of the target software system. The drawback of this approach is that the original developers are typically not available.
\item
Documentation-Based Decomposition: The decomposition is obtained  using the key functional concepts extracted from the software architecture documentation. 
\item
Maintenance Log-Based Decomposition: The decomposition is obtained by extracting information embedded in maintenance logs, which can be utilized to produce multiple decomposition stages of the target system.
\end{itemize}

\begin{figure}
\centering
\includegraphics[width=3.4 in]                {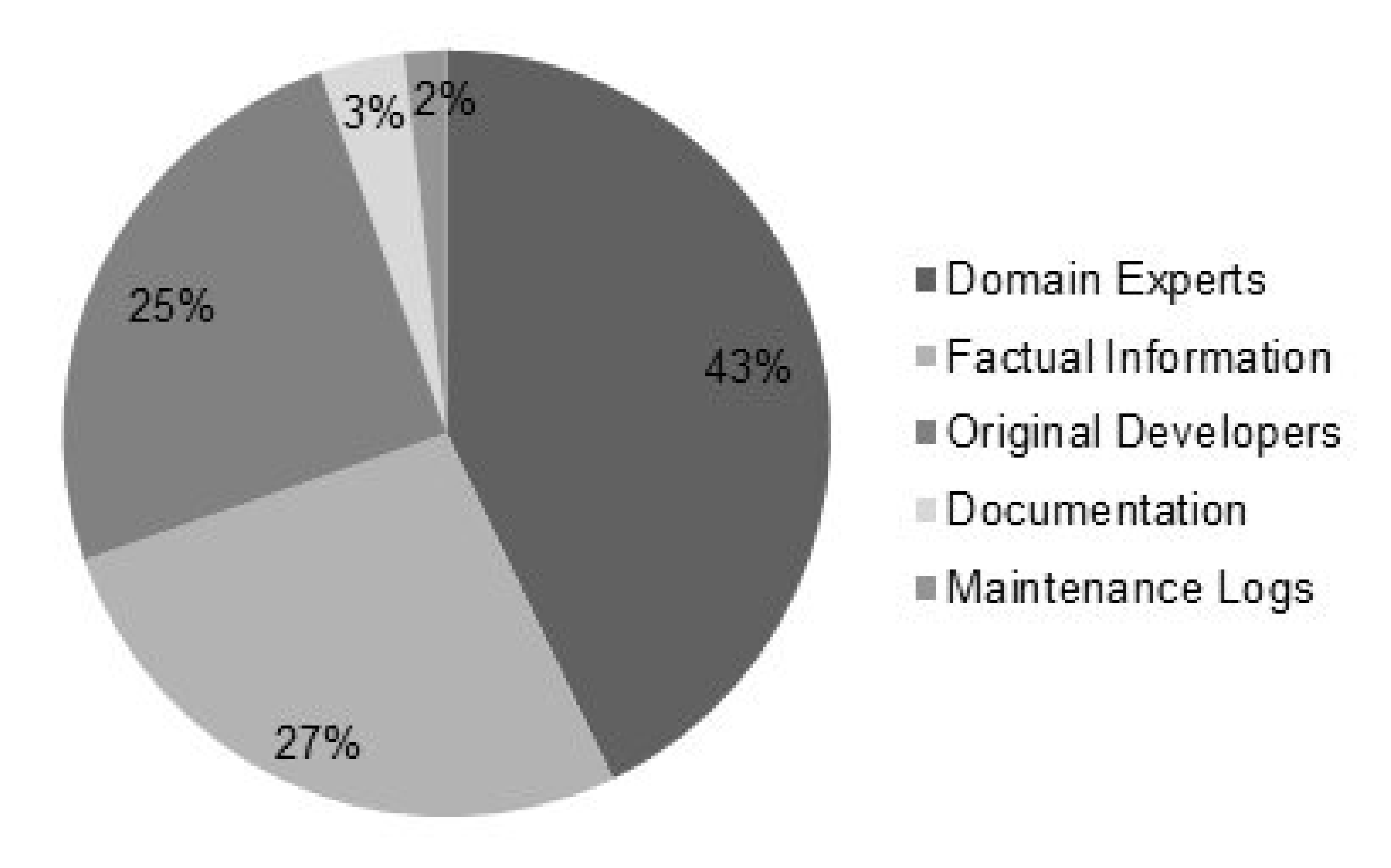}
\caption{Publication ratio by source of expert decomposition.}
\label{Fig:18}
\end{figure}

Many clustering approaches use expert decomposition to evaluate the clustering results. Figure \ref{Fig:18} shows the distribution of the expert decomposition, where 24 approaches are based on domain experts, 15 are based on factual information, 14 are based on the original developers, two are based on the documentation, and one is based on maintenance logs.

Many factors affect the quality and efficacy of a clustering process. The following points summarize those factors \cite{ref4-2010}:

\begin{itemize}
\item
The domain, type, size, and architecture of the targeted software system. For example, a clustering algorithm that is successful for a procedural program or a small software system might be unsuccessful for a large system developed in an object-oriented paradigm  \cite{ref10-2011}.
\item
The choice of factbase sources, preprocessing of data extracted from them and finding an appropriate representation of them \cite{ref3-2016}.
\item
The entities, features, and relationships between them in terms of how well they have been selected.
\item
The type of similarity measures and algorithms used for the clustering process \cite{ref3-2016}.
\item
Arbitrary decisions during the clustering process influence the quality and performance of clustering \cite{ref7-2010}. 
\end{itemize}


\subsection{Potential Future Directions of Research (RQ11)}

Answering RQ11 will help to identify possible research areas that may require further investigation. Based on the analysis of the considered papers, several potential directions of research were identified. The following points summarize and categorize these future research directions:

\begin{itemize}
\item

Scalability: The clustering approach should handle a growing quantity of input without decreasing the clustering results' quality. This process can be achieved by performing clustering in parallel using multithreading programming techniques and hardware systems that have multicore processors. The consistency of clustering must be ensured, i.e., performing the clustering multiple times on the same dataset should produce the same results.

\item
Visualization: The display of the clustering results should be improved in cases where large outputs are presented on the screen simultaneously. This improvement can be achieved by applying filters to separate the results into different abstraction layers. As a result, clusters within a specific layer only can be viewed instead of going through all the displayed results. Thus, the user can better understand the results. The visualization tools should also automatically associate labels with the generated clusters.

\item
CASE tool: It is essential to consider making the whole clustering process a working tool (e.g., third-party libraries, plugins, and standalone applications) available for researchers, software engineers, and practitioners to perform further experimentation and collect feedback for future improvements.

\item
Targeted systems: The targeted or subject software systems used in the experiments are mostly Java-based open-source systems. It would be interesting to examine the use of non-open-source systems and systems written in other popular programming languages. Furthermore, considering different software application domains to keep the results generic and widely applicable requires further study. When the targeted systems are selected from various application domains, a specific clustering algorithm may exhibit various performance characteristics. Therefore, the various performance features of a particular clustering algorithm, when applied to multiple target systems, should be investigated.

\item
Entity features: Some software entities such as files and classes have a large number of different types of features (e.g., lines of code, executable statements, number of functions, and number of variables or objects) to be extracted. From a practical perspective, addressing this large number of various features is not preferable. Therefore, experimental studies to determine the number and type of features needed for better clustering results should be conducted. Further experiments may reveal a clustering approach that is more suitable for target systems that present specific characteristics (e.g., size or implemented functionality).

\item
Factbase sources: There are different sources for factbase extraction. Each has its own set of features and drawbacks. Thus, experimental studies that can determine the type of factbase sources that provide better clustering results should be performed. The impact of integrating different factbase sources on the overall clustering accuracy should be studied.

\item
Cooperative clustering: Only a few studies have combined more than one clustering algorithm to achieve a common goal. Currently, clusters that are produced by the first clustering algorithm are subdivided and reclustered by the second clustering algorithm, a situation that is often undesirable. Thus, more experiments should be performed, and tools to address the situation by not reclustering all the clusters that are part of the initial solution set should be developed.

\item
Selection of clustering algorithms: The selection of appropriate software clustering algorithms plays a significant role in producing meaningful clustering results. The authors in \cite {other8} proposed guidelines for selecting or rejecting a clustering algorithm for a given software system. However, there are no comprehensive methods for clustering algorithm selection. Thus, further research and experiments can be conducted to provide formal selection methods based on empirical evidence.

\item 
Clustering with aesthetic aspects of the software design: The computational determination of the optimized cluster is often mechanistic, ignoring the fact that software is a creative artifact. Typically, cluster arrangement is determined via the grouping of nodes from dependency graphs. Having an optimized clustering of modules versus a meaningful set of clusters are two sides of the same coin with competing objectives. On the one hand, it is desirable to maximize cohesion and minimize coupling at any cost. On the other hand, one may also need to capture semantic as well as the essence and aesthetic aspects of the software design. Integrating natural language processing with deep learning and considering on other criteria (apart from simply grouping nodes from dependency graphs), such as utilizing the defined naming of modules and internal variables or other factbase sources as part of clustering arrangement criteria, would be a useful endeavor.

\item
Beyond clustering: The ultimate aim of software module clustering is to help software engineers apply the recommended clustering results to their software projects. However, do software engineers really adhere to the  recommended clustering results? If yes, what will be the impact, cost, or effort of realizing the suggested results? No research has comprehensively addressed these issues. Thus, a thorough investigation in this respect may be a good step for further study.

\end{itemize}


\section{Threats to validity}\label{Threats}
Every literature mapping study has a number of threats that might affect its validity. In this study, several threats were eliminated by considering well-known recommendations and guidelines on conducting literature mapping studies as follows:

\begin{itemize}
\item
Coverage of research questions: The threat here is that the research questions of this study may not cover all the aspects of the state-of-the-art research in software module clustering. To address this threat, all the authors of this study used brainstorming to define the desired set of research questions that cover the existing research in the area. 

\item
Coverage of relevant studies: It cannot be guaranteed that  all the relevant studies in software module clustering have been identified. Accordingly, different literature databases have been used, and a PICO method-based search string with various term synonyms (each author of the paper suggested different terms that lead to desired clustering concepts) has been applied to obtain the relevant research publications. However, some unidentified papers may remain. To address this issue, the snowballing method was intensively applied to reduce the possibility of missing important related papers.

\item
Paper inclusion/exclusion criteria: Application of the criteria can suffer from single-author judgment and personal bias. To address this issue, each paper was included or excluded for this study only after the authors reached a consensus.

\item
Accuracy of data extraction: Data extraction can suffer from the single-author experience. Accordingly, each author individually performed the data extraction process, and the outcomes of all authors were compared in an online meeting. In the meeting, all authors discussed differences between the outcomes until a final and agreed consensus was reached. Automatic filtering provided by Microsoft Excel was also used to ensure the accuracy of the data extraction process.

\item
Reproducibility of the study: The issue here is whether other researchers can perform this study with similar results. Accordingly, all the steps followed and performed in this study were reported in the research methodology (see Section \ref{ResearchMethod}).

\end{itemize}

\section{Conclusion}\label{Conclusion}

This paper systematically reports the state-of-the-art empirical contributions in software module clustering. Thus, to ascertain the recent clustering applications in software engineering, the algorithms and tools used to enable the software module clustering process were identified. A total of 143 papers from popular literature databases published in the area of software module clustering from 2008-2019 were selected for this study. The published papers were a combination of works from conferences, journals, symposiums, and workshops. However, most of the published papers were from conferences. From different perspectives and based on several identified RQs, the selected studies were thoroughly reviewed and analyzed. The findings were in different categories. For instance, statistics on the published studies, their publication venues, active authors, and countries were reported. Then, software module clustering applications were categorized. All the algorithms, tools, target software systems, evaluations, and metrics that enabled the clustering process were briefly discussed. Finally, as there are many research studies on software module clustering, novice researchers are likely to experience difficulties in addressing different aspects of the area. Therefore, we propose this analysis study as a primary reference to simplify the process of finding the most relevant information.

\section{ACKNOWLEDGEMENT}

This work has been partially funded by the Knowledge Foundation of Sweden (KKS) through the Synergy Project AIDA - A Holistic AI-driven Networking and Processing Framework for Industrial IoT (Rek:20200067).

\bibliographystyle{IEEEtran}
\bibliography{sample.bib}


\clearpage
\onecolumn

{\scriptsize
\begin{longtable}{ |l|l|p{14.7cm}|l| }
  
\caption{List of all papers included in the study.}
\label{Tab:11}

\\
 \hline

ID\#  & Ref. & Paper Title  & Year  \\ 
\hline
1   & \cite{ref1-2008} & A Double K-Clustering Approach for restructuring Distributed Object-Oriented software                                                  & 2008  \\ 
\hline
2   & \cite{ref2-2008} & An algorithm of system decomposition based on laplace spectral graph partitioning technology                                           & 2008  \\ 
\hline

3   & \cite{ref4-2008} & Assessing software archives with evolutionary clusters                                                                                 & 2008  \\ 
\hline
4   & \cite{ref5-2008} & Cluster analysis of Java dependency graphs                                                                                             & 2008  \\ 
\hline
5   & \cite{ref6-2008} & Clustering based automatic refactorings identification                                                                                 & 2008  \\ 
\hline
6   & \cite{ref7-2008} & Employing Clustering for Assisting Source Code Maintainability Evaluation according to ISO / IEC- 9126                                 & 2008  \\ 
\hline
7   & \cite{ref8-2008} & Evolution Strategy Based Automated Software Clustering Approach                                                                        & 2008  \\ 
\hline

8  & \cite{ref11-2008} & Object-Oriented Software Systems Restructuring through Clustering                     & 2008  \\ 
\hline

9  & \cite{ref13-2008} & 
Refactoring module structure                     & 2008  \\ 

\hline
10  & \cite{ref12-2008} & Using Cluster Analysis to Improve the Design of Component Interfaces                                                                   & 2008  \\ 
\hline
11  & \cite{ref1-2009} & An Approach for Software Architecture Refactoring Based on Clustering of Extended Component Dependency Graph                           & 2009  \\ 
\hline
12  & \cite{ref2-2009} & Clustering for Monitoring Software Systems Maintainability Evolution                                                                   & 2009  \\ 
\hline
13  & \cite{ref3-2009} & Clustering of Software Systems Using New Hybrid Algorithms                                                                             & 2009  \\ 
\hline
14  & \cite{ref4-2009} & Comparison of Graph Clustering Algorithms for Recovering Software Architecture Module Views                                            & 2009  \\ 
\hline
15  & \cite{ref5-2009} & Decomposing object-oriented class modules using an agglomerative clustering technique                                                  & 2009  \\ 
\hline
16  & \cite{ref6-2009} & Design pattern directed clustering for understanding open source code                                                                  & 2009  \\ 
\hline
17  & \cite{ref7-2009} & Restructuring Distributed Object-Oriented Software Using Hierarchical Clustering                                                       & 2009  \\ 
\hline
18  & \cite{ref8-2009} & Software Clustering Using Dynamic Analysis and Static Dependencies                                                                     & 2009  \\ 
\hline
19  & \cite{ref9-2009} & Splitting a large software repository for easing future software evolution-an industrial experience report                             & 2009  \\ 
\hline
20  & \cite{ref10-2009} & Towards automating class-splitting using betweenness clustering                                                                        & 2009  \\ 
\hline

21  & \cite{ref2-2010} & A Density Based Clustering approach for early detection of fault prone modules                                                         & 2010  \\ 
\hline
22  & \cite{ref3-2010} & A Probabilistic Based Approach towards Software System Clustering                                                                      & 2010  \\ 
\hline
23  & \cite{ref4-2010} & Architecture Recovery Using Latent Semantic Indexing and K-Means: An Empirical Evaluation                                              & 2010  \\ 
\hline
24  & \cite{ref5-2010} & Evaluating the Impact of Software Evolution on Software Clustering                                                                     & 2010  \\ 
\hline
25  & \cite{ref6-2010} & Hierarchical clustering for adaptive refactorings identification                                                                       & 2010  \\ 
\hline
26  & \cite{ref7-2010} & Improved Hierarchical Clustering Algorithm for Software Architecture Recovery                                                          & 2010  \\ 
\hline
27  & \cite{ref8-2010} & Object-oriented software architecture recovery using a new hybrid clustering algorithm                                                 & 2010  \\ 
\hline
28  & \cite{ref9-2010} & On the Comparability of Software Clustering Algorithms                                                                                 & 2010  \\ 
\hline

29  & \cite{ref11-2010} & Software architecture reconstruction: An approach based on combining graph clustering and partitioning                                 & 2010  \\ 
\hline
30  & \cite{ref12-2010} & Software refactoring at the function level using new Adaptive K-Nearest Neighbor algorithm                                             & 2010  \\ 
\hline
31  & \cite{ref13-2010} & Using the Kleinberg Algorithm and Vector Space Model for Software System Clustering                                                    & 2010  \\ 
\hline
32  & \cite{ref1-2011} & Applying agglomerative hierarchical clustering algorithms to component identification for legacy systems                               & 2011  \\ 
\hline
33  & \cite{ref2-2011} & Assessing Software Quality by Program Clustering and Defect Prediction                                                                 & 2011  \\ 
\hline
34  & \cite{ref3-2011} & Clustering and lexical information support for the recovery of design pattern in source code                                           & 2011  \\ 
\hline
35  & \cite{ref4-2011} & Clustering Dynamic Class Coupling Data to Measure Class Reusability Pattern                                                            & 2011  \\ 
\hline
36  & \cite{ref5-2011} & Clustering software systems to identify subsystem structures using knowledgebase                                                       & 2011  \\ 
\hline
37  & \cite{ref6-2011} & Clustering Support for Static Concept Location in Source Code                                                                          & 2011  \\ 
\hline
38  & \cite{ref7-2011} & Deriving High-level Abstractions from Legacy Software Using Example-driven Clustering                                                  & 2011  \\ 
\hline
39  & \cite{ref8-2011} & Investigating the Use of Lexical Information for Software System Clustering                                                            & 2011  \\ 
\hline
40  & \cite{ref9-2011} & JDeodorant: Identification and Application of Extract Class Refactorings                                                               & 2011  \\ 
\hline
41  & \cite{ref10-2011} & Object oriented software clustering based on community structure                                                                       & 2011  \\ 
\hline
42  & \cite{ref11-2011} & Software Module Clustering as a Multi-Objective Search Problem                                                                         & 2011  \\ 
\hline
43  & \cite{ref12-2011} & Software refactoring at the package level using clustering techniques                                                                  & 2011  \\ 
\hline
44  & \cite{ref13-2011} & Solving software module clustering problem by evolutionary algorithms                                                                  & 2011  \\ 
\hline
45  & \cite{ref9-2012} & An analysis of the effects of composite objectives in multiobjective software module clustering                                                 & 2012  \\ 
\hline
46  & \cite{ref1-2012} & Clustering Source Code Files to Predict Change Propagation during Software Maintenance                                                 & 2012  \\ 
\hline
47  & \cite{ref2-2012} & Comparing and Combining Genetic and Clustering Algorithms for Software Component Identification from Object-Oriented Code              & 2012  \\ 

\hline
48  & \cite{ref8-2012} & Evaluating relationship categories for clustering object-oriented software systems              & 2012  \\ 

\hline
49  & \cite{ref3-2012} & Feature-gathering dependency-based software clustering using Dedication and Modularity                                                 & 2012  \\ 
\hline
50  & \cite{ref4-2012} & Maintenance activities in object oriented software systems using K-means clustering technique: A review                                & 2012  \\ 
\hline
51  & \cite{ref5-2012} & Program restructuring using agglomerative clustering technique based on binary features                                                & 2012  \\

\hline

52  &  \cite{ref7-2012} & Reconstructing Architectural Views from Legacy Systems                                                & 2012  \\ 
\hline
53  & \cite{ref6-2012} & Software Clustering: Unifying Syntactic and Semantic Features                                                                          & 2012  \\ 
\hline
54& \cite{ref10-2012} & Towards module-based automatic partitioning of Java applications                                                                & 2012  \\ 
\hline
55  & \cite{ref11-2012} & Using fold-in and fold-out in the architecture recovery of software systems                                                                & 2012  \\ 
\hline
56  & \cite{ref1-2013} & A new hierarchical clustering technique for restructuring software at the function level                                               & 2013  \\ 
\hline
57  & \cite{ref2-2013} & Class level fault prediction using software clustering                                                                                 & 2013  \\ 
\hline
58  & \cite{ref3-2013} & Clustering Software Components for Component Reuse and Program Restructuring                                                           & 2013  \\ 
\hline
59  & \cite{ref4-2013} & Cooperative clustering for software modularization                                                                                     & 2013  \\ 
\hline
60  & \cite{ref5-2013} & Document Clustering Using Hybrid XOR Similarity Function for Efficient Software Component Reuse                                        & 2013  \\ 
\hline
61  & \cite{ref6-2013} & Efficient software clustering technique using an adaptive and preventive dendrogram cutting approach                                   & 2013  \\ 
\hline
62  & \cite{ref7-2013} & Evaluating software clustering algorithms in the context of program comprehension                                                      & 2013  \\ 
\hline
63  & \cite{ref8-2013} & Mixed-Integer Linear Programming Formulations for the Software Clustering Problem                                                      & 2013  \\ 
\hline
64  & \cite{ref9-2013} & Software architecture decomposition using adaptive K-nearest neighbor algorithm                                                        & 2013  \\ 
\hline
65  & \cite{ref10-2013} & Software Architecture Decomposition Using Clustering Techniques                                                                        & 2013  \\ 
\hline
66  & \cite{ref11-2013} & Software Clustering Using Automated Feature Subset Selection                                                                           & 2013  \\ 
\hline
67  & \cite{ref12-2013} & Software module clustering using a hyper-heuristic based multi-objective genetic algorithm                                             & 2013  \\ 
\hline
68  & \cite{ref13-2013} & Using spectral clustering to automate identification and optimization of component structures                                          & 2013  \\ 
\hline
69  & \cite{ref11-2014} & A clustering-based model for class responsibility assignment problem in object-oriented analysis                                & 2014  \\
\hline
70  & \cite{ref1-2014} & A Package Based Clustering for enhancing software defect prediction accuracy                                                           & 2014  \\ 
\hline
71  & \cite{ref2-2014} & An empirical study of the sensitivity of quality indicator for software module clustering                                              & 2014  \\ 
\hline
72  & \cite{ref9-2014} & 
Assessing modularity using co-change clusters                                              & 2014  \\ 
\hline
73  & \cite{ref10-2014} & 
Combining Clustering and Classification for Software Quality Evaluation                                              & 2014  \\ 
\hline

74  & \cite{ref5-2014} & Cooperative based software clustering on dependency graphs                                                                             & 2014  \\ 

\hline

75  & \cite{ref12-2014} & High dimensional search-based software engineering: Finding Tradeoffs Among 15 Objectives for Automating Software Refactoring Using NSGA-III                                                                    & 2014  \\ 
\hline
76  & \cite{ref6-2014} & Remodularization analysis using semantic clustering                                                                                    & 2014  \\ 
\hline

77  & \cite{ref8-2014} & Software modularization using the modified firefly algorithm                                                                           & 2014  \\ 
\hline
78  & \cite{ref1-2015} & A clustering technique based on the specifications of software components                                                              & 2015  \\ 
\hline
79  & \cite{ref2-2015} & A search-based approach to multi-view clustering of software systems                                                                   & 2015  \\ 
\hline
80  & \cite{ref3-2015} & Adaptive Clustering Techniques for Software Components and Architecture                                                                & 2015  \\ 
\hline
81  & \cite{ref4-2015} & CCIC: Clustering analysis classes to identify software components                                                                      & 2015  \\ 
\hline
82  & \cite{ref5-2015} & Clustering Source Code Elements by Semantic Similarity Using Wikipedia     & 2015  \\ 

\hline
83  & \cite{ref12-2015} & 
Constrained agglomerative hierarchical software clustering with hard and soft constraints     & 2015  \\

\hline

84  & \cite{ref7-2015} & Multi-objective Module Clustering for Kate                                                                                             & 2015  \\ 
\hline
85  & \cite{ref8-2015} & Object oriented based technique for software quality prediction through clustering and chi-square test                                 & 2015  \\ 
\hline
86  & \cite{ref9-2015} & Software Architecture Recovery using Genetic Black Hole Algorithm                                                                      & 2015  \\ 
\hline
87  & \cite{ref10-2015} & Software Clone Detection Using Clustering Approach                                                                                     & 2015  \\ 
\hline
88  & \cite{ref11-2015} & Source Code Driven Enterprise Application Decomposition: Preliminary Evaluation                                                        & 2015  \\ 
\hline
89  & \cite{ref1-2016} & A New Binary Similarity Measure Based on Integration of the Strengths of Existing Measures: Application to Software Clustering         & 2016  \\ 
\hline
90  & \cite{ref2-2016} & A similarity-based modularization quality measure for software module clustering problems                                              & 2016  \\ 
\hline
91  & \cite{ref3-2016} & A software component selection technique based on fuzzy clustering                                                                     & 2016  \\ 
\hline
92  & \cite{ref4-2016} & A tool to support software clustering using the software evolution information                                                         & 2016  \\ 
\hline
93  & \cite{ref5-2016} & Automatic clustering of code changes  & 2016  \\ 

\hline
94  & \cite{ref13-2016} & 
Clones clustering using K-means  & 2016  \\

\hline
95  & \cite{ref6-2016} & Clustering Software Metric Values Extracted from C\# Code for Maintainability Assessment                                               & 2016  \\ 
\hline

96  & \cite{ref8-2016} & Hyper-heuristic Approach for Multi-Objective Software Module Clustering                                                                & 2016  \\ 
\hline
97  & \cite{ref9-2016} & Implementation and evaluation of optimized algorithm for software architectures analysis through unsupervised learning (clustering)    & 2016  \\ 
\hline

98  & \cite{ref11-2016} & Modularizing Software Systems using PSO optimized hierarchical clustering& 2016  \\ 
\hline

99  & \cite{ref14-2016} &  Software Evolution Information Driven Service-Oriented Software Clustering & 2016  \\ 

\hline
100  & \cite{ref12-2016} & Weighing lexical information for software clustering in the context of architecture recovery                                           & 2016  \\ 

\hline
101  & \cite{ref11-2017} & A hierarchical clustering-based approach for software restructuring at the package level   & 2017  \\ 

\hline
102  & \cite{ref1-2017} & A multi-agent evolutionary algorithm for software module clustering problems                                                           & 2017  \\ 
\hline
103 & \cite{ref2-2017} & A Particle Swarm Optimization-Based Heuristic for Software Module Clustering Problem                                                   & 2017  \\ 
\hline
104 & \cite{ref9-2017} & 
Class Modularization Using Indirect Relationships                        & 2017  \\ 
\hline
105 & \cite{ref3-2017} & FP-ABC: Fuzzy-Pareto dominance driven artificial bee colony algorithm for many-objective software module clustering                    & 2017  \\ 

\hline
106 & \cite{ref12-2017} & Framework Information Based Java Software Architecture Recovery                    & 2017  \\ 

\hline
107 & \cite{ref13-2017} &  Improved binary similarity measures for software modularization                  & 2017  \\ 

\hline
108 & \cite{ref10-2017} &  Improving package structure of object-oriented software using multi-objective optimization and weighted class connections                  & 2017  \\ 
\hline
109 & \cite{ref4-2017} & Large Neighborhood Search applied to the Software Module Clustering problem                                                            & 2017  \\ 
\hline
110 & \cite{ref5-2017} & On the significance of relationship directions in clustering algorithms for reverse engineering                                        & 2017  \\ 

\hline
111 & \cite{ref14-2017} & 
Reconstructing and evolving software architectures using a coordinated clustering framework                                        & 2017  \\ 

\hline
112 & \cite{ref6-2017} & Semantic-based software clustering using hill climbing                                                                                 & 2017  \\ 
\hline
113 & \cite{ref7-2017} & Software Remodularization by Estimating Structural and Conceptual Relations Among Classes and Using Hierarchical Clustering            & 2017  \\ 
\hline
114 & \cite{ref8-2017} & Using hierarchical agglomerative clustering to locate potential aspect interference                                                    & 2017  \\ 
\hline
115 & \cite{ref1-2018} &  A design structure matrix approach for measuring co-change-modularity of software products                                             & 2018  \\ 
\hline

116 & \cite{ref3-2018} & Analyzing the structure of Java software systems by weighted K-core decomposition                                                      & 2018  \\ 
\hline
117 & \cite{ref4-2018} & Automatic Software Refactoring via Weighted Clustering in Method-Level Networks                                                        & 2018  \\ 
\hline
118 & \cite{ref5-2018} & BCD: Decomposing Binary Code Into Components Using Graph-Based Clustering                                                              & 2018  \\ 
\hline

119 & \cite{ref7-2018} & Discovering Program Topoi via Hierarchical Agglomerative Clustering                                                                    & 2018  \\ 
\hline
120 & \cite{ref8-2018} & Effectively incorporating expert knowledge in automated software remodularisation                                                      & 2018  \\ 
\hline
121 & \cite{ref9-2018} & Functionality-Oriented Microservice Extraction Based on Execution Trace Clustering                                                     & 2018  \\ 
\hline
122 & \cite{ref10-2018} & Improving Cohesion of a Software System by Performing Usage Pattern Based Clustering                                                   & 2018  \\ 
\hline
123 & \cite{ref11-2018} & Improving Problem Identification via Automated Log Clustering using Dimensionality Reduction                                           & 2018  \\ 
\hline
124 & \cite{ref12-2018} & Improving reusability of software libraries through usage pattern mining                                                               & 2018  \\ 
\hline
125 & \cite{ref13-2018} & In Object-Oriented Software Framework Improving Maintenance Exercises Through K-Means Clustering Approach                              & 2018  \\ 
\hline

126 & \cite{ref14-2018} & Many-objective artificial bee colony algorithm for large-scale software module clustering problem                                      & 2018  \\ 

\hline
127 & \cite{ref17-2018} & Software Module Clustering Algorithm Using Probability Selection & 2018  \\

\hline
128 & \cite{ref18-2018} & Software Module Clustering Based on the Fuzzy Adaptive Teaching Learning Based Optimization Algorithm & 2018  \\

\hline
129 & \cite{ref1-2019} & A Mechanism for Automatically Summarizing Software Functionality from Source Code  
 & 2019  \\
 
 \hline
130 & \cite{ref15-2019} &   A multi-objective search based approach to identify reusable software components
 & 2019  \\

\hline
131 & \cite{ref2-2019} &  A new algorithm for software clustering considering the knowledge of dependency between artifacts in the source code  
 & 2019  \\

\hline
132 & \cite{ref3-2019} &  A novel approach for automatic remodularization of software systems using extended ant colony optimization algorithm 
 & 2019  \\

\hline
133 & \cite{ref4-2019} &   An efficient and stable method to cluster software modules using ant colony optimization algorithm 
 & 2019  \\

\hline
134 & \cite{ref5-2019} &  Code similarity detection through control statement and program features 
 & 2019  \\
 
 \hline
135 & \cite{ref6-2019} &   Euclidean space based hierarchical clusterers combinations: an application to software clustering 
 & 2019  \\

\hline
136 & \cite{ref7-2019} &  Evaluating the effectiveness of multi-level greedy modularity clustering for software architecture recovery 
 & 2019  \\
 
 \hline
137 & \cite{ref8-2019} &  GUI-based software modularization through module clustering in edge computing based IoT environments 
 & 2019  \\
 
 \hline
138 & \cite{ref9-2019} &   Hybrid of genetic algorithm and krill herd for software clustering problem 
 & 2019  \\
 
  \hline
139 & \cite{ref10-2019} &  Multi-programming language software systems modularization 
 & 2019  \\
 
 \hline
140 & \cite{ref11-2019} &  Software Architecture Module-View Recovery Using Cluster Ensembles 
 & 2019  \\
 
 \hline
141 & \cite{ref12-2019} &  Software clusterings with vector semantics and the call graph
 & 2019  \\
 
 \hline
142 & \cite{ref13-2019} &  Software Modularization by Combining Genetic and Hierarchical Algorithms
 & 2019  \\
 
 \hline
143 & \cite{ref14-2019} &  Tarimliq: A new internal metric for software clustering analysis
 & 2019  \\

\hline

\end{longtable}
}

\clearpage
\twocolumn


\clearpage
\onecolumn

\clearpage
\onecolumn

\begin{table*}[h]
    \centering
\begin{longtable}{ |l|p{11cm}| } 
\caption{List of active journals with abbreviations.}\label{Tab:12}\\
\hline
Acronym                              & Journal Full Name                                                     \\ 
\hline

Inf. Softw. Technol.               & Information and Software Technology                                   \\ 
\hline

J. Syst. Softw.                    & Journal of Systems and Software                                       \\ 
\hline

IEEE Trans. Softw. Eng.            & IEEE Transactions on Software Engineering                             \\
\hline

Procedia Comput. Sci.                & Procedia Computer Science                                             \\ 
\hline

Soft Comput.                         & Soft Computing                                                        \\ 
\hline

IET Softw.                         & IET Software                                                          \\ 
\hline

J. Software Maint. Evol. Res. Pract. & Journal of Software Maintenance and Evolution: Research and Practice  \\ 
\hline

Inf. Sci.                            & Information Sciences                                                  \\ 
\hline

Wuhan Univ. J. Nat. Sci.             & Wuhan University Journal of Natural Sciences                          \\ 
\hline

IEEE Trans. Reliab.                  & IEEE Transactions on Reliability~                                     \\ 
\hline

Future Gener. Comput. Syst.          & Future Generation Computer Systems                                    \\ 
\hline

Empir. Softw. Eng.                   & Empirical Software Engineering                                        \\ 
\hline

Comput. Oper. Res.                   & Computers and Operations Research                                     \\ 
\hline

Expert Syst. Appl.           &        Expert Systems with Applications                      \\
 \hline

J. Supercomput.            &        Journal of Supercomputing                     \\
\hline

 Comput. Lang. Syst. Struct.          & Computer Languages, Systems, and Structures~~~~~~~~~~~~               \\ 
\hline

Comput. Optim. Appl.                 & Computational Optimization and Applications                           \\ 
\hline

Arabian J. Sci. Eng.             & Arabian Journal for Science and Engineering                           \\ 
\hline

Adv. Eng. Softw.                   & Advances in Engineering Software                                      \\
\hline

J. King Saud Univ. Comp. Info. Sci.                   & Journal of King Saud University: Computer and information sciences                                      \\ 
\hline

Front. Comput. Sci.      & Frontiers of Computer Science                                \\ 
\hline

Form. Asp. Comput.      & Formal Aspects of Computing                               \\ 
\hline

Cluster Comput.           &        Cluster Computing                      \\
 \hline

 J. Ambient Intell. Humaniz. Comput.           &        Journal of Ambient Intelligence and Humanized Computing
                     \\
\hline

Comput. Electr. Eng.  &  Computers and Electrical Engineering
                     \\
 \hline
 
  J. Comput. Lang. &  Journal of Computer Languages
                     \\  
\hline

IEEE Access  &  IEEE Access
                     \\  
\hline

Front. Inf. Technol. Electron. Eng.  &  Frontiers of Information Technology and Electronic Engineering
                     \\     
\hline

Autom. Softw. Eng.  &  Automated Software Engineering \\              
\hline

ACM SIGSOFT Software Eng. Notes      & ACM SIGSOFT Software Engineering Notes                                \\ 
\hline

\end{longtable}
\end{table*}

\clearpage
\twocolumn

\clearpage
\twocolumn

\clearpage
\onecolumn

\begin{table*}[h]
 \centering   
  
\begin{longtable}{ |l|p{14.6cm}| } 
\caption{List of active conferences with abbreviations.}\label{Tab:13}\\
\hline
Acronym & Conference Full Name                                                          \\ 
\hline

ICPC    & International Conference on Program Comprehension (ICPC)                      \\
\hline

CSMR    & European Conference on Software Maintenance and Reengineering (CSMR)          \\ 
\hline

WCRE    & Working Conference on Reverse Engineering (WCRE)                              \\ 
\hline

ASE     & International Conference on Automated Software Engineering (ASE)              \\ 
\hline

ICSM    & International Conference on Software Maintenance (ICSM)                       \\ 
\hline

CSSE    & International Conference on Computer Science and Software Engineering (CSSE)  \\ 
\hline

MySEC   & Malaysian Software Engineering Conference (MySEC)                             \\ 
\hline

CCECE   & Canadian Conference on Electrical and Computer Engineering (CCECE)            \\ 
\hline

COMPSAC & Annual Computer Software and Applications Conference (COMPSAC)                \\ 
\hline

SETN    & Hellenic Conference on Artificial Intelligence (SETN)        \\
\hline

GECCO    & International Conference on Genetic and Evolutionary Computation (GECCO)                      \\
\hline

KBEI    & International Conference on Knowledge Based Engineering and Innovation (KBEI)        \\
\hline

\end{longtable}
\end{table*}

\clearpage
\twocolumn

\end{document}